\documentclass{elsart}
\usepackage{natbib}
\usepackage{epsfig}
\usepackage{amssymb}
\newtheorem{approks}[thm]{Approximation}

\begin{document}

\begin{frontmatter}

\title{Heuristic average-case analysis of the backtrack resolution 
of random 3-Satisfiability instances.}

\author{Simona Cocco}
\address{CNRS-Laboratoire de Dynamique des Fluides Complexes,
3 rue de l'universit\'e, 67000 Strasbourg, France}
\ead{cocco@ldfc.u-strasbg.fr}
\ead[url]{http://ludfc39.u-strasbg.fr/cocco/index.html}

\author{R\'emi Monasson}
\address{CNRS-Laboratoire de Physique Th{\'e}orique de l'ENS,
24 rue Lhomond, 75005 Paris, France;\\
CNRS-Laboratoire de Physique Th{\'e}orique, 3 rue de l'universit\'e, 
67000 Strasbourg, France}
\ead{monasson@lpt.ens.fr}
\ead[url]{http://www.lptens.fr/$\sim$monasson}

\begin{abstract}
An analysis of the average-case 
complexity of solving random 3-Satisfiability (SAT)
instances with backtrack algorithms is presented. We first interpret
previous rigorous works in a unifying framework based on the statistical
physics notions of dynamical trajectories, phase diagram and growth
process. It is argued that, under
the action of the Davis--Putnam--Loveland--Logemann (DPLL) algorithm,
3-SAT instances are turned into $2+p$-SAT instances whose characteristic
parameters (ratio $\alpha$ of clauses per variable, fraction $p$ of
3-clauses) can be followed during the operation, and define resolution
trajectories. Depending on the location of trajectories in the
phase diagram of the 2+p-SAT model, easy (polynomial) or hard
(exponential) resolutions are generated.  Three regimes are identified,
depending on the ratio $\alpha$ of the 3-SAT instance to be
solved. Lower sat phase: for small ratios, DPLL almost surely finds a
solution in a time growing linearly with the number $N$ of
variables. Upper sat phase: for intermediate ratios, instances are
almost surely satisfiable but finding a solution requires exponential
time ($\sim 2 ^{N\,\omega}$ with $\omega>0$) 
with high probability.  Unsat phase: for
large ratios, there is almost always no solution and proofs of
refutation are exponential. An analysis of the growth of the search 
tree in both upper sat and unsat regimes is presented, and allows us
to estimate $\omega$ as a function of $\alpha$. This analysis is based
on an exact relationship between the average size of the search tree
and the powers of the evolution operator encoding the elementary 
steps of the search heuristic.  
\end{abstract}

\begin{keyword}
satisfiability, analysis of algorithms, backtrack.


\end{keyword}

\end{frontmatter}

\section{Introduction.}

This paper focuses on the average complexity of solving random 3-SAT
instances using backtrack algorithms.  Being an NP-complete problem,
3-SAT is not thought to be solvable in an efficient way, {\em i.e.}
in time growing at most polynomially with $N$. In practice, one
therefore resorts to methods that need, {\em a priori}, exponentially
large computational resources. One of these algorithms is the
ubiquitous Davis--Putnam--Loveland--Logemann (DPLL) solving
procedure\citep{DP,survey}. DPLL is a complete search algorithm based on
backtracking; its operation is briefly recalled in Figure~1. 
The sequence of assignments of variables made by DPLL
in the course of instance solving can be represented as a search tree,
whose size $Q$ (number of nodes) is a convenient measure of the
hardness of resolution. 
Some examples of search trees are presented in Figure~\ref{trees}.

In the past few years, much experimental and theoretical progress
has been made on the probabilistic analysis of 3-SAT \citep{AI,Hans}.
Distributions of random instances controlled by few parameters are
particularly useful in shedding light on the onset of complexity. An
example that has attracted a lot of attention over the past years is
random 3-SAT: all clauses are drawn randomly and each variable negated
or left unchanged with equal probabilities. Experiments 
\citep{AI,Cra,Mit,Kir} and theory \citep{Friedgut,Dub,Dubtcs} indicate
that clauses can almost surely always (respectively never) be
simultaneously satisfied if $\alpha$ is smaller (resp. larger) than a
critical threshold $\alpha _C \simeq 4.3$ as soon as the numbers $M$
of clauses and $N$ of variables go to
infinity at a fixed ratio $\alpha$. This phase
transition \citep{Sta1} is accompanied by a drastic peak in
hardness at threshold \citep{AI,Mit,Cra}. The emerging pattern of
complexity is as follows.  At small ratios $\alpha < \alpha _L$, where
$\alpha _L$ depends on the heuristic used by DPLL, instances are
almost surely satisfiable (sat), see \citet{Fra2} and \citet{Achl} 
for recent
reviews.  The size $Q$ of the associated search tree scales, with high
probability, linearly with the number $N$ of variables, and almost no
backtracking is present \citep{Fri} (Figure~\ref{trees}A). Above the
critical ratio, that is when $\alpha >\alpha _C$, instances are
a.s. unsatisfiable (unsat) and proofs of refutation are obtained
through massive backtracking (Figure~\ref{trees}B), leading to an
exponential hardness: $Q = 2^{N\omega}$ with $\omega >0$~\citep{Chv}.
In the intermediate range, $\alpha _L <\alpha <\alpha _C$, finding a
solution a.s. requires exponential effort ($\omega >0$)
\citep{Vardi1,Achl3,Coc}. 

The aim of this article is two-fold. First, we propose a simple and
intuitive framework to unify the above findings. This framework is
presented in Section 2. It is based on the statistical physics notions
of dynamical trajectories and phase diagram, and was, to some extent,
implicitly contained in the pioneering analysis of search heuristics
by \citet{fra2,Fra}.  Secondly, we present in Section 3 a quantitative
study of the growth of the search tree in the unsat regime. Such a
study has been lacking so far due to the formidable difficulty in
taking into account the effect of massive backtracking on the
operation of DPLL. We first establish an exact relationship 
between the average size of the search tree and the powers
of the evolution operator encoding the elementary steps of the
search heuristic. This equivalence is then used (in a non rigorous way) to
accurately estimate the logarithm $\omega$ of the average complexity
$Q$ as a function of $\alpha$,
\begin{equation}
\omega (\alpha) = \lim _{N\to \infty}\; \frac 1N\; 
\log _2 \; E_{(\alpha,N)} [Q] \quad ,
\end{equation}
where $E_{(N,\alpha)}$ denotes the expectation value for given $N$
and $\alpha$. The approach emphasizes the relevance of
partial differential equations to analyse algorithms in presence of
massive backtracking, as opposed to ordinary differential equations in
the absence of the latter \citep{Worm,Achl}. In Section 4, we focus upon
the upper sat regime {\em i.e.} upon ratios $\alpha _L < \alpha <
\alpha_C$.  Combining the framework of Section 2 and the analysis of
Section 3 we unveil the structure of the search tree
(Figure~\ref{trees}C) and calculate $\omega$ as a function of the
ratio $\alpha$ of the 3-SAT instance to be solved.
 
For the sake of clarity and since the style of our approach may look
unusual to the computer scientist reader, the status of the different
calculations and results (experimental, exact, conjectured,
approximate, ...) are made explicit throughout the article.

\begin{figure}
\begin{center}
\includegraphics[width=280pt,angle=-0]{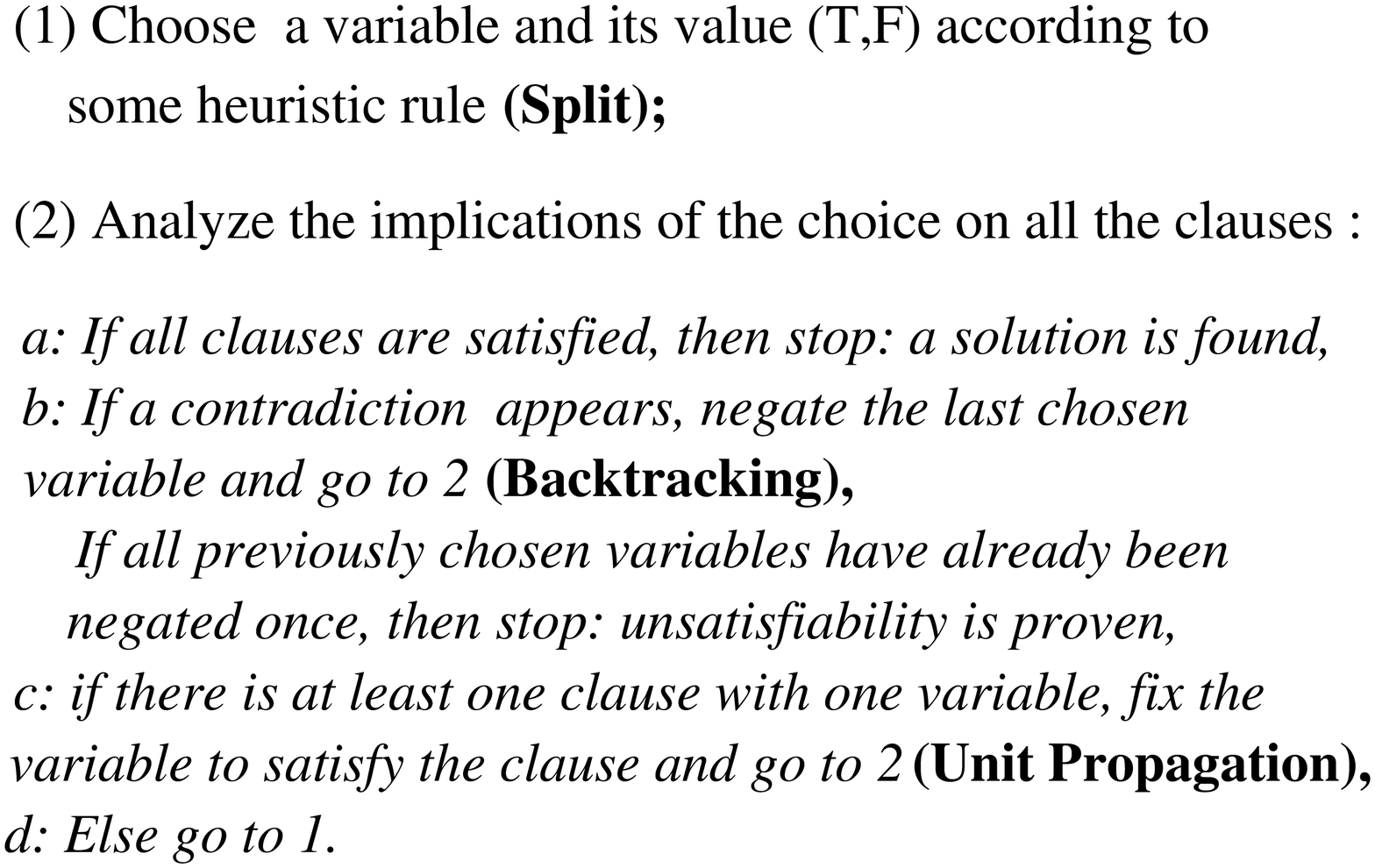}
\vskip .5cm
\caption{DPLL algorithm.  When a variable has been chosen at step (1)
e.g. $x=T$, at step (2) some clauses are satisfied e.g. $C=(x \;\hbox
{\rm OR}\;{y} \;\hbox {\rm OR}\; z)$ and eliminated, other are reduced
e.g. $C=(\hbox{\rm not}\, x\; \hbox {\rm OR}\;{y}\; \hbox {\rm OR}\;
{z}) \to C=({y}\; \hbox {\rm OR}\; {z})$. If some clauses include one
variable only e.g. $C=y$, the corresponding variable is automatically
fixed to satisfy the clause ($y=T$).  This 
propagation (2c) is repeated up to the exhaustion of all
unit clauses. Contradictions result from the presence of two opposite 
unit clauses
e.g. $C=(y), C'=(\hbox{\rm not}\, y)$. A solution is found when no clauses
are left.  The search process of DPLL is represented by a tree 
(Figure~\ref{trees}) whose
nodes correspond to (1), and edges to (2).  Branch extremities are
marked with contradictions C (2B,2C), or by a solution S (2A,2C).}
\end{center}
\end{figure}

\begin{figure}
\begin{center}
\includegraphics[width=190pt,angle=-90]{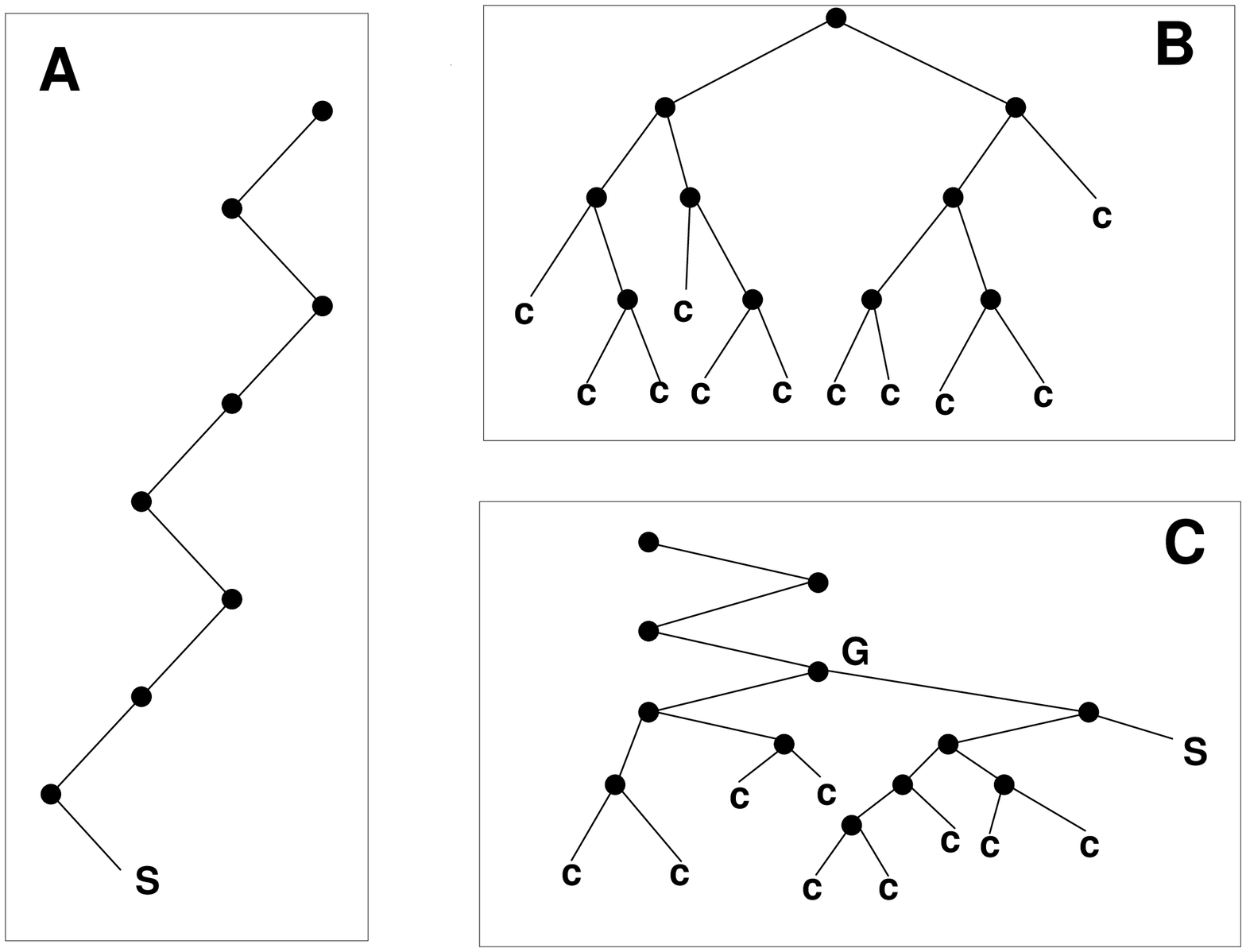}
\end{center}
\vskip .5cm
\caption{Types of search trees generated by the DPLL solving procedure
on random 3-SAT. 
{\bf A.} {\em simple branch:} the algorithm finds
easily a solution without ever backtracking. {\bf B.} {\em dense tree:}
in the absence of solution, DPLL builds a tree, including many branches 
ending with contradictory leaves, before stopping.  {\bf C.} {\em
mixed case, branch + tree:} if many contradictions arise before
reaching a solution, the resulting search tree can be decomposed into a
single branch followed by a dense tree. G is the highest
node in the tree reached by DPLL through backtracking.}
\label{trees}
\end{figure}

\section{Phase diagram and trajectories.}

\subsection{The 2+p-SAT distribution and split heuristics}

\begin{center}
\begin{figure}
\includegraphics[width=300pt,angle=-90]{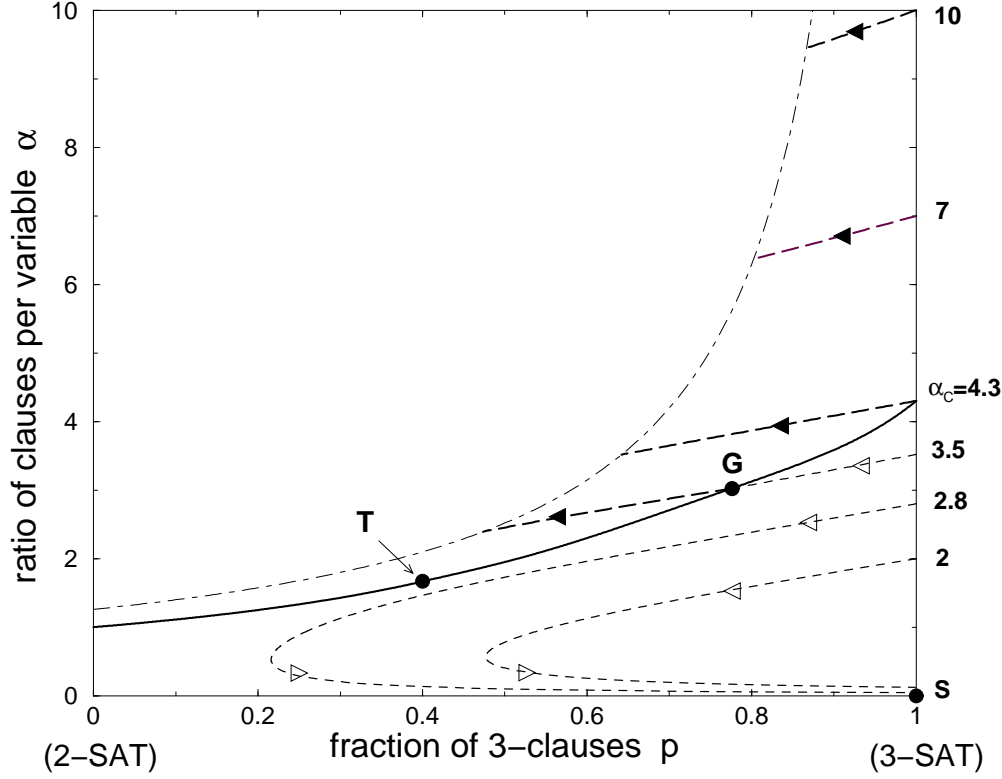}
\vskip 1cm
\caption{Phase diagram of 2+p-SAT and dynamical trajectories of DPLL.
The threshold line $\alpha_C (p)$ (bold full line) separates sat
(lower part of the plane) from unsat (upper part) phases. Extremities
lie on the vertical 2-SAT (left) and 3-SAT (right) axis at coordinates
($p=0,\alpha _C=1$) and ($p=1,\alpha _C\simeq 4.3$) respectively.  
Departure points for DPLL trajectories are located on the
3-SAT vertical axis and the corresponding values of $\alpha$ are
explicitely given. Dashed curves represent tree trajectories in the
unsat region (thick lines, black arrows) and branch
trajectories in the sat phase (thin lines, empty
arrows). Arrows indicate the direction of "motion" along trajectories
parametrized by the fraction $t$ of variables set by DPLL.  For small
ratios $\alpha < \alpha _L$, branch trajectories remain
confined in the sat phase, end in S of coordinates $(1,0)$, where a
solution is found. At $\alpha_L$ ($\simeq 3.003$ for the GUC heuristic), 
the single branch trajectory hits
tangentially the threshold line in T of coordinates $(2/5,5/3)$. In
the intermediate range $\alpha _L < \alpha < \alpha_C$, the branch
trajectory intersects the threshold line at some point G (which depends
on $\alpha$). A dense tree then grows in the unsat phase, as happens
when 3-SAT departure ratios are above threshold $\alpha > \alpha _C
\simeq 4.3$. The tree trajectory halts on the dot-dashed curve
$\alpha \simeq 1.259/(1-p)$ where the tree growth process stops.
At this point, DPLL has reached back the highest backtracking node in
the search tree, that is, the first node when $\alpha > \alpha _C$, or node G
for $\alpha _L < \alpha < \alpha_C$.
In the latter case, a solution can be reached from a new descending branch
while, in the former case, unsatisfiability is proven, see Figure~2.}
\label{diag}
\end{figure}
\end{center}

The action of DPLL on an instance of 3-SAT causes changes to the
overall numbers of variables and clauses, and thus of the ratio
$\alpha$.  Furthermore, DPLL reduces some 3-clauses to 2-clauses.  A
mixed 2+p-SAT distribution, where $p$ is the fraction of 3-clauses,
can be used to model what remains of the input instance at a node of
the search tree. Using experiments and methods from statistical
mechanics \citep{Sta1}, the threshold line $\alpha _C (p)$, separating
sat from unsat phases, may be estimated with the results shown in
Figure~\ref{diag}. For $p \le p_0 = 2/5$, {\em i.e.} to the left of
point T, the threshold line is given by $\alpha _C(p)=1/(1-p)$, as
rigorously confirmed by \citet{Achl1}, and saturates the upper bound
for the satisfaction of 2-clauses. Above $p_0$, no exact value for
$\alpha _C (p)$ is known. Note that $\alpha _C \simeq 4.3$ corresponds to
$p=1$.

The phase diagram of 2+p-SAT is the natural space in which DPLL
dynamic takes place. An input 3-SAT instance with ratio $\alpha$ shows
up on the right vertical boundary of Figure~\ref{diag} as a point of
coordinates $(p=1,\alpha )$.  Under the action of DPLL, the
representative point moves aside from the 3-SAT axis and follows a
trajectory. This trajectory obviously depends on the heuristic of split 
followed by DPLL (Figure~1). 
Possible simple heuristics are \citep{fra2,Fra},
\begin{itemize}
\item{\em Unit-Clause (UC):} 
randomly pick up a literal among a unit clause 
if any, or any unset variable otherwise.

\item{\em Generalized Unit-Clause (GUC):}  
randomly pick up a literal  among the shortest avalaible clauses.

\item{\em Short Clause With Majority (SC$_1$):}  
randomly pick up a literal among unit clauses if any; otherwise
randomly pick up  
an unset variable $v$, count the numbers of occurences $\ell,
\bar \ell$ of $v$, $\bar v$ in 3-clauses, and choose $v$ (respectively
$\bar v$) if $\ell > \bar \ell$ (resp.  $\ell < \bar \ell$). When
$\ell=\bar \ell$, $v$ and $\bar v$ are equally likely to be chosen.
\end{itemize}

Rigorous mathematical analysis, undertaken to provide rigorous bounds to
the critical threshold $\alpha _C$,  have so far been restricted to
the action of DPLL prior to any backtracking, that is, to the
first descent of the algorithm in the search tree\footnote{The analysis of
\cite{Fri} however includes a very limited version of backtracking, see
Section 2.2}. 
The corresponding search branch is drawn on Figure~\ref{trees}A. 
These studies rely on the two following facts:

First, the representative point
of the instance treated by DPLL does not ``leave'' the 2+p-SAT phase
diagram. In other words, the instance is, at any stage of the search 
process, uniformly distributed from the 2+p-SAT 
distribution conditioned to its clause--per--variable ratio $\alpha$ and 
fraction of 3-clauses $p$. This assumption is not true for
all heuristics of split, but holds for the above examples ($UC$, $GUC$, 
$SC_1$) \citep{fra2}. Analysis of more sophisticated heuristics 
require to handle more complex instance distributions \citep{Kap}.

Secondly, the trajectory followed by an instance in the course of
resolution is a stochastic object, due to the randomness of 
the instance and of the assignments done by DPLL. 
In the large size limit ($N\to\infty$), this trajectory gets 
concentrated around its average locus in the 2+p-SAT phase diagram. 
This concentration phenomenon results from general properties of 
Markov chains \citep{Worm, Achl}.

\subsection{Trajectories associated to search branches}

Let us briefly recall \citet{fra2} analysis of the
average trajectory corresponding to the action of DPLL prior to
backtracking. The ratio of clauses per variable of the 
3-SAT instance to be solved will be denoted by $\alpha _0$.
The numbers of 2 and 3-clauses are initially equal to $C_2=0, C_3=\alpha _0
\, N$ respectively. Under the action of DPLL, $C_2$ and $C_3$ follow 
a Markovian stochastic evolution process, as the depth $T$ along the branch 
(number of assigned variables) increases. Both $C_2$ and $C_3$ are 
concentrated around their expectation values, the densities 
$c_j (t)= E[C_j( T=t\,N)]$ ($j=2,3$) of which obey a set of 
coupled ordinary differential equations (ODE) \citep{fra2,Fra,Achl},
\begin{equation}
\frac{d c_3}{dt} = - \frac{ 3\, c_3}{1-t} \qquad , \qquad
\frac{d c_2}{dt} = \frac{ 3\, c_3}{2(1-t)} - \frac{ 2\, c_2}{1-t} -
\rho _1 (t) \; h(t) \qquad , \label{ode}
\end{equation}
where $\rho _1 (t) = 1 - c_2(t)/(1-t)$ is the probability that DPLL fixes a 
variable  at depth $t$ (fraction of assigned variables) 
through unit-propagation. Function $h$ depends upon the
heuristic: $h_{UC} (t)=0$, $h_{GUC} (t)=1$ (if $\alpha _0> 2/3$;
for $\alpha _0<2/3$, see \citet{Fra}), 
$h_{SC_1}
(t)=a\, e^{-a}\, (I_0(a)+I_1(a))/2$  where $a\equiv 3\, c_3(t)/(1-t)$
and $I_\ell$ is the $\ell^{th}$ modified Bessel function.
To obtain the single branch trajectory in the phase diagram of 
Figure~\ref{diag},
we solve ODEs (\ref{ode}) with initial conditions $c_2(0)=0, 
c_3(0)=\alpha_0$, and perform the change of variables
\begin{equation}
p(t) = \frac{c_3(t)}{c_2(t)+c_3(t)} \qquad , \qquad
\alpha (t) = \frac{c_2(t)+c_3(t)}{1-t} \quad . \label{change}
\end{equation}

Results are shown for the GUC heuristics and starting ratios $\alpha_0 =2$
and 2.8 in Figure~\ref{diag}. The trajectory,
indicated by a light dashed line, first heads to the left and then
reverses to the right until reaching a point on the 3-SAT axis at
a small ratio. Further action of
DPLL leads to a rapid elimination of the remaining clauses and the
trajectory ends up at the right lower corner S, where a solution is
found.

\citet{Fri} have shown that, for ratios $\alpha _0 < \alpha _L \simeq 3.003$
(for the GUC heuristics), the full search tree essentially reduces
to a single branch, and is thus entirely described by the ODEs (\ref{ode}).
The amount of backtracking necessary to reach a solution is bounded from above 
by a power of $\log N$. The average size of the branch, $Q$, scales linearly
with $N$ with a multiplicative factor $\gamma (\alpha _0)=Q/N$ that can
be calculated \citep{Coc}.
The boundary $\alpha _L$ of this easy sat region can be defined as the largest
initial ratio $\alpha _0$ such that the branch trajectory $(p(t),\alpha (t))$ 
issued from $(1,\alpha _0)$ never leaves the sat phase during DPLL action.
In other words, the instance essentially keeps being sat throughout the
resolution process. We shall see in Section 4 this does not hold  for sat 
instances with ratios $\alpha _L < \alpha _0 < \alpha _C$. 

\section{Analysis of the search tree growth in the unsat phase.}

In this Section, we present an analysis of search trees
corresponding to unsat instances, that is, in presence of massive 
backtracking. We first report results from numerical experiments,
then expose our analytical approach to compute the complexity of
resolution (size of search tree). 

\subsection{Numerical experiments}

\begin{center}
\begin{figure}
\hskip 1.5cm \includegraphics[height=250pt,angle=-90]{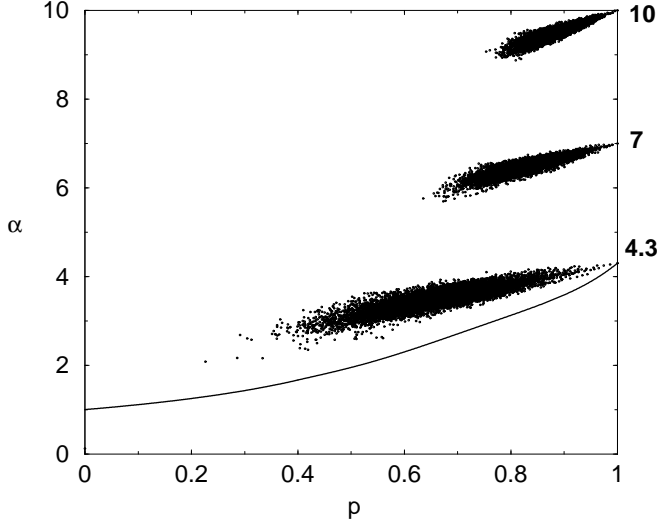}
\caption{Clouds associated to search trees obtained from the resolution 
of three unsat
instances with initial ratios $\alpha _0= 4.3, 7$ and 10
respectively. Each point in the cloud corresponds to a splitting node
in the search tree. Sizes of instances and search trees are
$N=120,Q=7597$ for $\alpha _0=4.3$, $N=200,Q=6335$ for $\alpha _0 =7$,
and $N=300,Q=6610$ for $\alpha _0=10$.}
\label{patch}
\end{figure}
\end{center}

For ratios above threshold ($\alpha _0 > \alpha _C\simeq 4.3$),
instances almost never have a solution but a considerable amount of
backtracking is necessary before proving that clauses are
incompatible. Figure~\ref{trees}B shows a generic unsat, or
refutation, tree.  In contrast to the previous section, the sequence
of points $(p,\alpha)$ attached to the nodes of the search tree 
do not arrange along a line any longer, but
rather form a cloud with a finite extension in the phase diagram of
Figure~\ref{diag}.  Examples of clouds are provided on
Figure~\ref{patch}.

\begin{table}
$$
\begin{array}{|c|c|c|c|c|c||c|}
\hline 
\multicolumn{1}{|c} {\alpha _0}& 
\multicolumn{1}{|c} {4.3} & \multicolumn{1}{|c} 7 & 
\multicolumn{1}{|c} {10} & \multicolumn{1}{|c} {15} &  
\multicolumn{1}{|c||} {20} & \multicolumn{1}{|c|} {3.5} \\
\hline
\omega _{EXP} &
0.089  & 0.0477  & 0.0320  &0.0207  & 0.0153 & 0.034 \\
& \pm 0.001  & \pm 0.0005 &\pm 0.0005 &
\pm 0.0002 & \pm 0.0002 & \pm 0.003   \\
\hline
\omega _{THE} & 0.0916 & 0.0486 & 0.0323 &
0.0207  & 0.0153 &  0.035  \\
\hline 
\end{array}
$$
\vskip .3cm
\caption{Logarithm of the complexity $\omega$ from experiments (EXP) and 
theory (THE) as a function of the ratio $\alpha _0$ of clauses per 
variable of the 3-SAT instance. 
Ratios above 4.3 correspond to unsat instances; the rightmost ratio
lies in the upper sat phase.}
\end{table}

The number of points in a cloud {\em i.e.} the size $Q$ of its
associated search tree grows exponentially with $N$ \citep{Chv}. It is
thus convenient to define its logarithm $\omega$ through $Q=2^{N
\omega}$. We experimentally measured $Q$, and averaged its 
logarithm $\omega$ over a large number of
instances. Results have then be extrapolated to the $N\to \infty$
limit \citep{Coc} and are reported in Table~1.  $\omega$ is a
decreasing function of $\alpha _0$ \citep{Bea}: the larger $\alpha
_0$, the larger the number of clauses affected by a split, and the
earlier a contradiction is detected.  We will use the vocable
``branch'' to denote a path in the refutation tree which joins the top
node (root) to a contradiction (leaf). The number of branches, $B$, is
related to the number of nodes, $Q$, through the relation $Q=B-1$
valid for any complete binary tree. As far as exponential (in $N$)
scalings are concerned, the logarithm of $B$ (divided by $N$) equals
$\omega$.  In the following paragraph, we show how $B$ can be
estimated through the use of a matrix formalism.

\begin{center}
\begin{figure}
\hskip 1.5cm \includegraphics[height=150pt,angle=0]{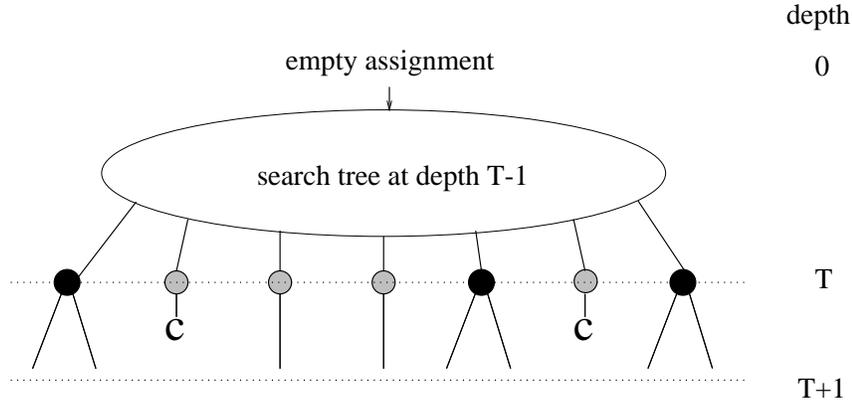}
\vskip .5cm
\caption{Imaginary, parallel growth process of an unsat search tree used in the
theoretical analysis. Variables are fixed through unit-propagation, or 
by the splitting heuristic as in the DPLL 
procedure, but branches evolve in parallel. $T$ denotes the depth in the
tree, that is the number of variables assigned by DPLL along each 
branch. At depth $T$, one literal is chosen on each branch among 1-clauses
(unit-propagation, grey circles not represented on Figure 2), or 2,3-clauses 
(splitting, black circles as in Figure~2).
If a contradiction occurs as a result of unit-propagation, the branch gets 
marked with C and dies out. The growth of the tree proceeds  
until all branches carry C leaves. The resulting tree is identical to the one
built through the usual, sequential operation of DPLL. }
\label{struct}
\end{figure}
\end{center}

\subsection{Parallel growth process and Markovian evolution matrix}

The probabilistic analysis of DPLL in the unsat regime appears to be a
formidable task since the search tree of Figure~\ref{trees}B is the
output of a complex, sequential process: nodes and edges are added by
DPLL through successive descents and backtrackings (depth-first
search). We have imagined a
different building up of the refutation tree, which results in the
same complete tree but can be mathematically analyzed. In our
imaginary process (Figure~\ref{struct}), the tree grows in parallel, 
layer after layer (breadth-first search).  
At time $T=0$, the tree reduces to a root
node, to which is attached the 3-SAT instance to be solved, and an
attached outgoing edge. At time $T$, that is, after having assigned
$T$ variables in the instance attached to each branch, the tree is
made of $B(T) \ (\le 2^T)$ branches, each one carrying a partial
assignment of variables. At
next time step $T\to T+1$, a new layer is added by assigning,
according to DPLL heuristic, one more variable along every branch. As a
result, a branch may keep growing through unitary propagation, get hit
by a contradiction and die out, or split if the partial assignment
does not induce unit clauses. This parallel growth process is
Markovian, and can be encoded in an instance--dependent matrix we now
construct.

To do so, we need some preliminary definitions:

\begin{defn} Partial state of variables.

The partial state $s$ of a Boolean variable $x$ is one of the three
following possibilities: undetermined ($u$) if the variable has not
been assigned by the search heuristic yet, true ($t$) if the variable
is partially assigned to true, false ($f$) if the variable is
partially assigned to false. The partial state $S$ of a set of Boolean
variables $X=\{x_1, x_2, \ldots , x_N\}$ is the collection of the
states of its elements, $S=\{s_1, s_2, \ldots , s_N \}$.

Let $\bf I$ be an instance of the SAT problem, defined over a set of
Boolean variables $X$ with partial state $S$. A clause of $\bf I$ is
said to be
\begin{itemize}
\item satisfied if at least one of its literals is true according to $S$;
\item unsatisfied, or violated if all its literals are false according to $S$;
\item undetermined otherwise; then its `type' is the number $(=1,2,3)$ of 
undetermined variables it includes. 
\end{itemize}
The instance $\bf I$ is said to be satisfied if all its clauses are 
satisfied, unsatisfied
if one (at least) of its clauses is violated, undetermined otherwise.
The set of partial states that violate ${\bf I}$ is denoted by $W$.  
\end{defn}
 
\begin{defn} Vector space attached to a variable.

To each Boolean variable $x$ is associated 
a three dimensional vector space ${\bf v}$ with spanning basis $|u\rangle$,  
$|t\rangle$, $|f\rangle$, orthonormal with respect
to the dot (inner) product denoted by $\langle . | . \rangle$,
\begin{equation}
\langle u | u \rangle = \langle t | t \rangle = \langle f | f \rangle = 
1\ ,\ \langle u | 
t \rangle = \langle u | f \rangle = \langle t | f \rangle = 0 \ .
\end{equation}
The partial state attached to a basis vector $|s\rangle$ is $s$ (=$\,u,t,f)$. 
\end{defn}

Letters $u$, $t$, $f$ stand for the different partial states the variable may 
acquire in the course of the search process. Note that the coefficients 
of the decomposition of any vector $|x\rangle \in {\bf v}$ 
over the spanning basis,
\begin{equation}
|x\rangle = x ^{(u)} \, |u\rangle + x^{(t)} \, |t\rangle  + x^{(f)} \, |f\rangle 
\quad ,
\end{equation}
can be obtained through use of the dot product: $x^{(s)} = \langle 
s | x \rangle$ with $s=u,t,f$. By extension, $\langle S|$ denotes the
transposed of vector $|S\rangle$.

\begin{defn} Vector space attached to a set of variables.

We associate to the set $X=\{x_1, x_2, \ldots , x_N\}$ of $N$ Boolean
variables the $3^N$--dimensional vector space ${\bf V}= {\bf v}_1
\otimes {\bf v}_2 \otimes \ldots \otimes {\bf v}_N$. The spanning
basis of ${\bf V}$ is the tensor product of the spanning basis of the
${\bf v}_i$'s. To lighten notations, we shall write $|s_1 , s_2, ...,
s _N \rangle$ for $|s _1\rangle \otimes |s _2 \rangle \otimes \ldots
\otimes |s _N\rangle$.  The partial state attached to a basis vector
$|S\rangle= |s_1 , s_2, ..., s _N \rangle$ is $S=(s_1 , s_2, ..., s
_N)$.  The dot product naturally extends over ${\bf V}$: $\langle s'_1
, s'_2 , \ldots , s'_N | s_1, s_2 , \ldots , s_N \rangle = 1$ if
$s_i=s' _i\ \forall i$, 0 otherwise.
\end{defn}

Any element $|X\rangle \in {\bf V}$ can be uniquely decomposed 
as a linear combination of vectors from the spanning basis.
Two examples of vectors are $|\Sigma\rangle$ and $|U\rangle$, 
respectively the sum of all vectors in the 
spanning basis and the fully undetermined vector,
\begin{eqnarray} \label{sigvec}
|\Sigma\rangle &=& \big( |u \rangle + |t \rangle + |f \rangle \big) \otimes
\big( |u \rangle + |t \rangle + |f \rangle \big) \otimes \ldots \otimes
\big( |u \rangle + |t \rangle + |f \rangle \big) \ , \\
|U\rangle &=& |u,u,\ldots , u\rangle \qquad . \label{uvec}
\end{eqnarray}
Basis vectors fulfill the closure identity
\begin{equation} \label{closure}
\sum _{S} | S\rangle \; \langle S | = {\bf 1} \quad ,
\end{equation}  
where ${\bf 1}$ is the identity operator on ${\bf V}$. To establish identity
(\ref{closure}), apply the left hand side
operator to any vector $|S'\rangle$ and take advantage of 
the orthonormality of the spanning basis .

\begin{defn} (Heuristic-induced) Transition probabilities

Let $S=(s_1 , s_2, ..., s _N)$ be a partial state which does not violate
instance ${\bf I}$.
Call $S^{(j,x)}$, with $j=1,\ldots, N$ and $x=t,f$, 
the partial state obtained from $S$ by replacing $s_j$ with $x$.
The probability that the heuristic under consideration (UC, GUC, ...) 
chooses to assign variable $x_j$ when presented partial state $S$ 
is denoted by $h(j|S)$.
The probability that the heuristic under consideration  
then fixes variable $x_j$ to $x\, (=\, t,f)$ is denoted by $g(x|S,j)$.
\end{defn}

A few elementary facts about  transition probabilities are:
\begin{enumerate}
\item $h(j|S)=0$ if $s_j\ne u$.
\item $g(x|S,j)+g(\bar x|S,j)=1$.
\item Assume that 
the number $C_1 (S)$ of undetermined clauses of type 1 (unit 
clauses) is larger or equal to unity. Call $C_1(j|S)$ the number of 
unit clauses containing variable $x_j$, and $C_1(x|S,j)$ the number
of unit clauses satisfied if $x_j$ equals
$x\,(=\,t,f)$. Clearly  $C_1(j|S)= C_1(t|S,j)+C_1(f|S,j)$.
Then, as a result of unit--propagation,
\begin{eqnarray}
h(j|S) &=&\frac{C_1(j|S)}{C_1(S)} \quad , \nonumber \\
g(x|S,j) &=& \frac{C_1 (x|S,j)}{C_1(j|S)} \quad 
\hbox{\rm for}\ x=t,f \
\hbox{\rm and} \ C_1(j|S) \ge 1 \ . 
\end{eqnarray}
\item In the absence of unitary clause ($C_1(S)=0$), transition probabilities depend on the details
of the heuristic. For instance, in the case of the UC heuristic,
\begin{enumerate}
\item if $s_j=u$, $h(j|S)=\frac{1}{u(S)}$ and $g(x|S,j)=\frac 12$,
\item if $s_j\ne u$, $h(j|S)=0$,
\end{enumerate}
where $u(S)$ is the number of undetermined variables in partial state $S$. 
\item The sum of transition probabilities from a partial state $S$
is equal to unity,
\begin{equation}
\sum _{j=1} ^N h (j|S) \; \bigg[ g(t|S,j) + g(f|S,j) \bigg] =1 \quad .
\end{equation}
\end{enumerate}
It is important to stress that the definition of the transition probabilities 
does not make any reference to any type of backtracking. It relies
on the notion of variable assignement through the heuristic of search only.

Let us now introduce the

\begin{defn} (Heuristic-induced) Evolution operator.

The evolution operator is a linear operator $\bf H$ acting on $\bf V$
encoding the action of DPLL for a given unsatisfiable instance ${\bf I}$. Its
matrix elements in the spanning basis are
\begin{enumerate}
\item if $S$ violates $\bf I$,
\begin{equation} \label{defmu}
\langle S'| {\bf H} | S\rangle
= \left\{ \begin{array} {c c}
1 & \hbox{\rm if} \ S'=S \\
0 & \hbox{\rm if} \ S' \ne S 
\end{array} \right. \qquad ,
\end{equation}
\item if $S$ does not violate $\bf I$,
\begin{equation} \label{defms}
\langle S'| {\bf H} | S\rangle
= \left\{ \begin{array} {c l}
h(j|S)\times g(x|S,j) & \hbox{\rm if} \ C_1(S) \ge 1  
\ \hbox{\rm and} \ S'=S^{(j,x)}\\
h(j |S) & \hbox{\rm if} \ C_1(S) =0\ \hbox{\rm and} \\
\ & \big(S'=S^{(j,x)}\ \hbox{\rm or} \ S'=S^{(j,\bar x)}\big) 
\\
0 & \hbox{\rm otherwise} 
\end{array} \right. \nonumber
\end{equation}
\end{enumerate}
where $S,S'$ are the attached partial states to $|S\rangle, |S'\rangle$,
and $C_1(S)$ is the number of undetermined clauses of type 1 (unitary clauses)
for partial state $S$.
\end{defn}

\begin{center}
\begin{figure}
\includegraphics[height=150pt,angle=0]{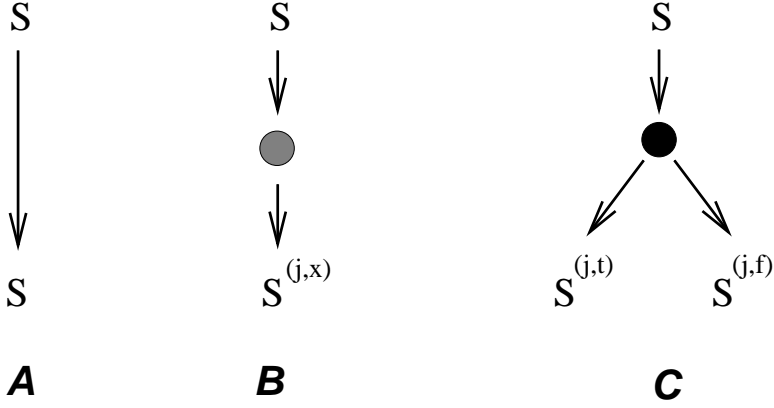}
\vskip .5cm
\caption{Transitions allowed by the heuristic-induced evolution
operator. Grey and black nodes correspond to variables assigned through 
unit-propagation and splitting respectively, as in Figure~\ref{struct}. 
{\bf A}. If the partial state $S$ already violates the instance 
${\bf I}$, it is left
unchanged. {\bf B}. If the partial state does not violate $\bf I$ and
there is at least one unitary clause, a variable is fixed through unit
propagation (grey node) e.g. $x_j=x$. The output partial state is $S^{j,x}$.
{\bf C}. If the partial state does not violate $\bf I$ and
there is no unitary clause, a variable $x_j$ is fixed through splitting
(black node). Two partial states are generated, $S^{j,t}$ and $S^{j,f}$.}
\label{thm}
\end{figure}
\end{center}

Notice that we use the same notation, $\bf H$, for the operator and its matrix
in the spanning basis. The different cases encountered in the above
definition of $\bf H$ are symbolized in Figure~\ref{thm}.
We may now conclude:

\begin{thm} Branch function and average size of refutation tree

Call branch function the function $B$ with integer-valued argument $T$,
\begin{equation} \label{funcq}
B(T) = \langle \Sigma | {\bf H}^T | U \rangle \quad ,
\end{equation}
where $\bf H$ is the evolution operator associated to the unsatisfiable 
instance $\bf I$, ${\bf H}^T$ denotes the $T^{th}$ (matricial) 
power of ${\bf H}$, and vectors $|\Sigma\rangle, |U\rangle$ are
defined in (\ref{sigvec},\ref{uvec}). 
Then, there exist two instance--dependent
integers $T^* \ (\le N)$ and $B^* \ (\le 2^N)$ such that,
\begin{equation} \label{statq}
B(T) = B^* \ , \quad \forall \ T \ge T^*  \ .
\end{equation}
Furthermore, $B^*$ is the expectation value over the random assignments
of variables of the size (number of leaves) of the search tree
produced by DPLL to refute ${\bf I}$. The smallest non zero $T^*$ for
which (\ref{statq}) holds is the largest number
of variables that the heuristic needs to assign to reach a contradiction. 
\end{thm}

\begin{pf*}{Proof of Theorem 6}

Let $S$ be a partial state. We call refutation tree built from
$S$ a complete search tree that proves the unsatisfiability of ${\bf I}$
conditioned to the fact that DPLL is allowed to assign only variables
which are undetermined in $S$. The height of the search tree is the
maximal number of assignments leading from the root node (attached to partial
state $S$) to a contradictory leaf.

Let $T$ be a positive integer. We call
$b_T (S)$ the average size (number of leaves) of refutation trees
of height $\le T$ that can be built from partial state $S$. Clearly,
$b_T(S)=1$ for all $S\in W$, and $b_T(S)\ge 2$ if $S\notin W$. Recall
$W$ is the set of violating partial states from Definition 1.

Assume now $T$ is an integer larger or equal to 1, $S$ a partial 
state with $C_1(S)$ unitary clauses. Our parallel representation of 
DPLL allows us to write simple recursion relations:
\begin{enumerate}
\item if $S \in W$, $b_T (S)=1=b_{T-1} (S)$. 
\item if $S \notin W$ and $C_1(S)\ge 1$,
\begin{equation}
b_T (S) = \sum _{j=1}^N \sum _{x=t,f} h(j |S)\, g(x |S,j) 
\; b_{T-1} \big(S ^{(j,x)} \big) \quad .
\end{equation}
\item if $S \notin W$ and $C_1(S)= 0$,
\begin{equation}
b_T (S) = \sum _{j=1}^N  h(j|S) \; \bigg[ 
b_{T-1} \big(S ^ {(j,t)} \big) + b_{T-1} \big(S ^ {(j,f)} \big) 
\bigg] \quad .
\end{equation}
\end{enumerate} 
These three different cases are symbolized on Figure~\ref{thm}A, B and
C respectively. From definitions (\ref{defmu},\ref{defms}), these
recursion relations are equivalent to
\begin{equation} \label{transpo}
b_T (S) = \sum _{S'} \langle S' | {\bf H} |S\rangle \;
b_{T-1} (S') \quad ,
\end{equation} 
for any partial state $S$. Let $|b_T\rangle$ be the vector of $\bf V$ whose
coefficients on the spanning basis $\{|S\rangle\}$ are the $b_T(S)$'s. 
In particular,
\begin{equation}
|b_0\rangle = \sum _{S_0 \in W} |S_0\rangle \quad .
\end{equation}
Then identity (\ref{transpo}) can be written as
$|b _T\rangle = {\bf H}^\dagger\; |b _{T-1} \rangle$
where ${\bf H}^\dagger$ is the transposed of the evolution operator.
Note that the branch function (\ref{funcq}) is simply $B(T)=\langle U|b_T
\rangle$. We deduce 
\begin{equation} \label{funcqprim}
|b _T\rangle = ({\bf H}^\dagger)^T\; |b_0 \rangle = \sum _{S_0\in W}
\sum _{\sigma _T}  p\big( \sigma _T;S_0\big) |S _ 0\rangle
\quad ,
\end{equation}
where the second sum runs over all $3^{N\times T}$ sequences $\sigma _T = 
(S_1, S_2 , \ldots , S_{T-1}, S _T)$ 
of $T$ partial states with associated weight
\begin{eqnarray}
p\big( \sigma _T;S_0\big) &=& \langle S_T |{\bf H}^\dagger 
|S_{T-1}\rangle \times  \ldots \times 
\langle S_2 |{\bf H}^\dagger|S_{1}\rangle \times \langle 
S _{1} |{\bf H}^\dagger|S _0\rangle \nonumber \\
 &=& \langle S _0 |{\bf H} |S_{1}\rangle 
\times  \langle S _1 |{\bf H} |S _{2}\rangle \ldots  
\times \langle S _{T-1} |{\bf H}|S _T\rangle
\ , \label{weightseq}
\end{eqnarray}
The length of a sequence is the number of partial states it includes. 
We call $S_0$--genuine a sequence of partial states $\sigma _T$ with 
non zero weight (\ref{weightseq}). The second sum 
on the right hand side of equation (\ref{funcqprim}) may be 
rewritten as a sum over all $S_0$--genuine sequences $\sigma _T$
of length $T$ only. 

\begin{lem}
Take $S_0\in W$. 
Any $S_0$-genuine sequence $\sigma _{N+1}$ of length $N+1$
includes at least one partial state belonging to $W$.  
\end{lem}
Suppose this is not true. There exists a genuine sequence
$\sigma _{{N+1}}$ with $S_{T}\notin W$, 
$\forall\ 1\le T\le N+1$. Call $u _T$ the number of undetermined variables in 
partial state $S_{T}$. Since the sequence is genuine, 
$\langle S_{T-1} |{\bf H} | S_{T}\rangle \ne 0$ 
for every $T$ comprised between 1 and $N+1$. From the evolution operator
definition (\ref{defms}), $S_{T}$ contains exactly one more 
undetermined variable than $S_{T-1}$, and $u _T=u _{T-1}+1$ for all 
$1\le T\le N+1$. Hence $u _{N+1}-u_0=N+1$. But 
$u_0$ and $u_{N+1}$ are, by definition, 
integer numbers comprised between 0 and $N$. 
\qed

From Lemma 7, the index $\nu$ of a $S_0$--genuine sequence $\sigma _{N+1}$
of length $N+1$,
\begin{equation}
\nu = \sup \big\{ T : 1\le T\le N+1\ \hbox{\rm and}\ 
S_{T} \in \sigma _{N+1} \ \hbox{\rm and}\ S_{T}\ \in \ W \big\} \quad ,
\end{equation}  
exists and is larger, or equal, to $1$. Let us define 
\begin{equation}
\hat \sigma _{N+1} = (S_{\nu +1}, S_{\nu+2}, 
\ldots, S_{N}, S_{N+1} )\quad .
\end{equation} 
From definition
(\ref{defmu}), $\sigma _{N+1}$ is simply $S_0$ repeated $\nu$ times 
followed by $\hat \sigma _{N+1}$, and $p (\sigma _{N+1})= p(\hat \sigma 
_{N+1})$. Call $\nu^*(S_0)$ the smallest index of all $S_0$-genuine
sequences of length $N+1$, and $\nu^*$ the minimum of $\nu^* (S_0)$ over 
$S_0\in W$.
Then, from equation (\ref{funcqprim}), $|b_{N+1}\rangle = |b_N\rangle =
\ldots = |b_{T^*}\rangle $
where $T^*=N+1-\nu^*\le N$. Thus $|b_{T^*}\rangle$ is a right eigenvector
of ${\bf H}^\dagger$ with eigenvalue unity, and
$|b _{T}\rangle= |b_{T^*}\rangle$ for all $T\ge T^*$.
$T^*$, which depends upon instance ${\bf I}$, 
is the length of the longest genuine sequence without repetition. 
It is the maximal
number of (undetermined) variables to be fixed before a contradiction is
found.

\begin{lem} Take $S\notin W$.
Then there is no $S$-genuine sequence of length $T^*$.
\end{lem}
Suppose this is not true. There exist $S\notin W$ and a $S$--genuine 
sequence $\sigma _{T^*}$ of length $T^*$. As $S$ does not violate ${\bf I}$,
and ${\bf I}$ is not satisfiable, there are still some undetermined
variables in partial state $S$. A certain number of them, say $T'\ge 1$,
must be assigned to some $t,f$ values to reach a contradiction, that is,
a partial state $S_0\in W$. Therefore there exists a $S_0$--genuine
sequence, $\tilde \sigma$, of length $T'\ge 1$ ending with $S$
and with no repeated partial state. 
Concatenating $\tilde \sigma$ and $\sigma _{T^*}$, we obtain a 
$S_0$--genuine sequence of length $T^*+T '>T^*$ and without repetition, 
in contradiction with the above result.
\qed

Using Lemma 8, we may replace $|b_0\rangle$ in equation (\ref{funcqprim})
with $|\Sigma\rangle$, and find
\begin{equation}
B(T)\equiv \langle \Sigma |{\bf H}^T |U\rangle = 
\langle U |({\bf H}^\dagger)^T |\Sigma\rangle =  \langle U | b_{T^*}\rangle
= b_{T^*} (U) \ ,
\end{equation}
for all $T\ge T^*$. Hence, $B^*=b_{T^*} (U)$ is the average size 
(over the random assignments made by the heuristic) of the refutation tree
to instance $\bf I$ generated from the fully undetermined partial state.
\qed
\end{pf*}

\subsection{Some examples of short instances and associated matrices}

\begin{center}
\begin{figure}
\includegraphics[height=150pt,angle=0]{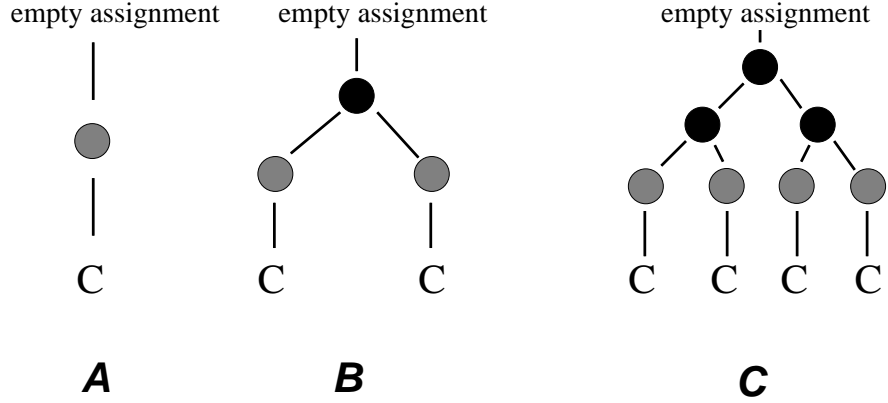}
\vskip .5cm
\caption{Refutation search trees associated to instances ${\bf I}_1$,
${\bf I}_2$ and ${\bf I}_3$. Grey and black nodes correspond to
variables assigned through unit-propagation and split respectively, as
in Figure~\ref{struct}.  {\bf A}. Example 9: refutation of instance
${\bf I}_1$ is obtained as a result of unit-propagation. The size
(number of leaves) of the search tree is $B=1$. {\bf B}. Example 10:
search tree generated by DPLL on instance ${\bf I}_2$. The black and
grey node correspond to the split of $x_1$ and unit-propagation over
$x_2$, or vice-versa.  The size of the tree is $B=2$. {\bf C}. Example
11: search tree corresponding to the instance ${\bf I}_3$ when DPLL
first splits variable $x_3$. The size of the tree is $B=4$. If the
first split variable is $x_1$ or $x_2$, the refutation search tree of
instance ${\bf I}_3$ corresponds to case {\bf B}.  }
\label{tree-exemple}
\end{figure}
\end{center}

We illustrate the above definitions and results with three explicit 
examples of instances involving few variables:

\begin{exmp}
Instance over $N=1$ variable

Consider the following unsat instance built from a single variable,
\begin{equation}
{\bf I}_1 = x_1 \wedge  \bar x_1 \qquad .
\end{equation}
The 3--dimensional vector space ${\bf v}_1$ is spanned by vectors $|u\rangle, |t\rangle, 
|f\rangle$. The evolution matrix reads
\begin{equation}
{\bf H} = \left( \begin{array}{c c c}
0 & 0 & 0 \\ \frac 12 & 1 & 0 \\  \frac 12 &  0 &1
\end{array} \right) \  \hbox{\rm with} \quad 
|u\rangle = \left( \begin{array}{c} 1 \\ 0 \\ 0 \end{array} \right) \ ,\quad 
|t\rangle = \left( \begin{array}{c} 0 \\ 1 \\ 0 \end{array} \right) \ ,\quad 
|f \rangle = \left( \begin{array}{c} 0 \\ 0 \\ 1 \end{array} \right) \ . 
\end{equation}
Entries can be interpreted as follows. 
Starting from the $u$ state, variable $x_1$ will be set through unit-propagation to $t$
or $f$ with equal probabilities: $\langle t |{\bf H }| 
u \rangle =\langle f |{\bf H} | u \rangle 
=1/2$. Once the variable has reached this state, the instance is violated: 
$\langle t |{\bf H} | t \rangle =\langle f |{\bf H} | f \rangle =1$. All other entries are null.  
In particular, state $u$ can never be reached from any state, 
so the first line of the matrix is filled in with zeroes: $\langle u |{\bf
H}| s \rangle =0,
\forall s$. Function (\ref{funcq}) is easily calculated
\begin{equation}
B (T)  =  \left( \begin{array}{c} 1 \\ 1 \\ 1 \end{array} \right) 
^\dagger \;.\; {\bf H}
^T\; . \; \left( \begin{array}{c} 1 \\ 0 \\ 0 \end{array} \right) =
1\ , \quad \forall\ T\ge 0 \ .
\end{equation}
Therefore, $T^*=B^*=1$. Indeed, refutation is obtained without
any split, and the search tree involves a unique branch of length 1 
(Figure~\ref{tree-exemple}A).  
\end{exmp}

Our next example is a 2-SAT instance whose refutation requires to split one 
variable.

\begin{exmp} Instance over $N=2$ variables, with a unique refutation tree.

\begin{equation}
{\bf I}_2 =(x_1 \vee x_2) \wedge (\bar x_1 \vee x_2) \wedge 
(x_1 \vee \bar x_2 ) \wedge (\bar  x_1 \vee \bar x_2)
\end{equation}
The evolution matrix $\bf H$ is a $9\times 9$ matrix with 16 non zero entries,
\begin{eqnarray}
\langle s,u | {\bf H} | u,u \rangle &=& 
\langle u,s | {\bf H} | u,u \rangle = \frac 12 \ , \qquad \forall\ s=t,f 
\label{m21} \\ \label{m22}
\langle s,s' | {\bf H} | s,u \rangle &=& 
\langle s,s' | {\bf H} | u,s' \rangle = \frac 12 \ , \qquad \forall\ s,s'=t,f\\
\label{m23}
\langle s',s | {\bf H} | s',s \rangle &=& 1\ , \qquad \forall \ s,s'=t,f \quad .
\end{eqnarray} 
We now explain how these matrix elements were obtained. From the
undetermined state $|u,u\rangle$, any of the four clause can be chosen
by the heuristic. Thus, any of the two literals $x_1$, $x_2$ has a
probability $1/2$ to be chosen: $h(1|u,u)=h(2|u,u)=\frac 12$. 
Next, unit-propagation will set the unassigned variable to true, or 
false with equal
probabilities $1/2$ (\ref{m22}). Finally, entries corresponding to
violating states in eqn (\ref{m23}) are calculated according to
rule (\ref{defmu}).

The branch function $B(T)$ equals 1 for $T=0$, 2 for any $T\ge 1$;
thus, $T^*=1$ and $B^*=2$, in agreement with the
associated search tree symbolized in Figure~\ref{tree-exemple}B.
\end{exmp}

We now introduce an instance with a non unique refutation tree.

\begin{exmp} Instance with $N=3$ variables, and two refutation trees.
\begin{equation}
{\bf I}_3 =(x_1 \vee x_2) \wedge (\bar x_1 \vee x_2) \wedge 
(x_1 \vee \bar x_2 ) \wedge (\bar  x_1 \vee \bar x_2) \wedge
(x_3 \vee \bar x_3)
\end{equation}
Notice the presence of a (trivial) clause containing opposite 
literals, which allows
us to obtain a variety in the search trees without considering more than three 
variables. The evolution matrix $\bf H$ is a $27\times 27$ matrix with 56 non 
zero entries (for the GUC heuristic),

\begin{eqnarray} \label{m31}
\langle s,u,u |{\bf H} | u,u,u \rangle &=& 
\langle u,s,u | {\bf H} | u,u,u \rangle = \frac 25  \ , \qquad \forall\ s=t,f \\ 
\label{m32}
\langle u,u,s | {\bf H} | u,u,u \rangle &=&  \frac 1{5} \ , \qquad \forall\ s=t,f \\
\langle s,s',s'' | {\bf H} | s,u,s'' \rangle &=& 
\langle s,s',s'' | {\bf H} | u,s',s'' \rangle = \frac 1{2} \ , \quad \forall\ s,s'=t,f;
s''=u,t,f \nonumber \\
\langle s',u,s | {\bf H} | u,u,s \rangle &=&
\langle u,s',s | {\bf H} | u,u,s \rangle = \frac 1{2} \ , \qquad \forall\ s,s'=t,f
\nonumber \\
\langle s,s',s'' | {\bf H} | s,s',s'' \rangle &=&1 \ , \qquad \forall\ s,s'=t,f;
\ s''=u,t,f \nonumber 
\end{eqnarray} 
The first split variable is $x_3$ if the last clause is chosen (probability $1/5$), 
or $x_1$ or $x_2$ otherwise (with probability $2/5$ each), leading to 
expressions (\ref{m31}) and (\ref{m32}). The remaining entries of $\bf H$ are 
obtained in the same way as explained in Example 10.

We obtain $B(0)=1$, $B(1)=2$ and $B(T\ge 2)=12/5$. 
Therefore, $T^*=2$ and 
\begin{equation}
B^* = \frac{12}5 = \frac 45 \times 2 + \frac 15 \times 4 \qquad ,
\end{equation}
where the different contributions to $B^*$ and their probabilities are explicitely
written down, see Figures~\ref{tree-exemple}B and \ref{tree-exemple}C. 
\end{exmp}

\subsection{Dynamical annealing approximation}

Let us denotes by $\overline q$ the expectation value of a
function $q$ of the instance ${\bf I}$ over the random 3-SAT distribution,
at given numbers of variable, $N$, and clauses, $\alpha \, N$.
From Theorem 6, the expectation value of the size of the refutation tree is
\begin{equation}
B^*(\alpha, N) \equiv \overline{ B^* }=
\langle \Sigma | \overline{ {\bf H} ^N} | U \rangle 
\quad .
\end{equation}
Calculation of the expectation value of the $N^{th}$ power of ${\bf H}$ is
a hard task that we were unable to perform for large sizes $N$. We therefore turned
to a simplifying approximation, hereafter called dynamical annealing.
This approximation is not thought to be justified in general, but may be 
asymptotically exact in some limiting cases we will expose later on. 

A first temptation is to approximate the expectation of the $N^{th}$ power of ${\bf H}$
with the $N^{th}$ power of the expectation of ${\bf H}$. This is however too a brutal
approximation to be meaningful, and a more refined scheme is needed.

\begin{defn} Clause projection operator

Consider an instance ${\bf I}$ of the 3-SAT problem. The clause vector
$\vec C (S)$ of a partial state $S$ is a three dimensional vector
$\vec C = (C_1, C_2, C_3)$ where $C_j$ is the number of undetermined
clauses of ${\bf I}$ of type $j$. The clause projection operator,
${\bf P} ({\vec C})$, is the operator acting on $\bf V$ and
projecting onto the subspace of partial state
vectors with clause vectors $\vec C$,
\begin{equation}
{\bf P} ({\vec C}) \; |S\rangle = \left[ \prod _{j=1}^3 \delta _{ C_j - C_j(S)}\right] \
|S\rangle \quad ,
\end{equation}
where $\delta$ is the Kronecker function. The sum of all state
vectors in the spanning basis with clause vector $\vec C$ is denoted by
$| \Sigma (\vec C) \rangle ={\bf P} ({\vec C})\; | \Sigma \rangle$.  
The sum of all state
vectors in the spanning basis with clause vector $\vec C$ and
$U$ undetermined variables is denoted by
$| \Sigma _U (\vec C) \rangle$.
\end{defn}

It is an easy check that ${\bf P}$ is indeed a projection operator:
${\bf P}^2({\vec C})={\bf P}({\vec C})$. 
As the set of partial states can be 
partitioned according to their clause vectors,
\begin{equation} \label{identriv}
\sum _{\vec C}  {\bf P} ({\vec C}) = \sum _{\vec C}  {\bf P}^2 ({\vec C})
={\bf 1} \quad .
\end{equation}
We now introduce the clause vector-dependent branch function 
\begin{equation}
B ( \vec C ,  T) = \langle \Sigma (\vec C) | {\bf H}^T |U \rangle
\quad .
\end{equation}
Summation of the $B$'s over all $\vec C$ gives back function (\ref{funcq})
from identity (\ref{identriv}).
The evolution equation for $B ( \vec C , T)$ is,
\begin{eqnarray} \label{ap2}
B ( \vec C , T+1)  &=& \langle \Sigma (\vec C)| {\bf H} \; 
\times {\bf H}^T |U\rangle 
\nonumber \\ 
&=& \langle \Sigma (\vec C)| {\bf H} \times   
\left( \sum _{\vec C'} {\bf P} ^2 ({\vec C'}) \right) \times {\bf H}^T |U\rangle 
\nonumber \\ 
&=& \sum _{\vec C'}\;  \langle \Sigma (\vec C)| {\bf H} \times   
 {\bf P}  ({\vec C'}) \times \left( \sum _S |S\rangle \langle S| 
\right)\times  {\bf P} ({\vec C'})     \times {\bf H}^T |U\rangle 
\nonumber \\ 
&=& \sum _{\vec C'} \sum _{S} \langle \Sigma (\vec C)|
{\bf H} \times  {\bf P} ({\vec C'}) | S\rangle \; \langle S|
 {\bf P} ({\vec C'})\times {\bf H}^T |U\rangle 
\end{eqnarray}
where we have made use of identities (\ref{closure}) and (\ref{identriv}).
We are now ready to do the two following approximation steps:

\begin{approks} Dynamical annealing (step A)

Substitute in equation (\ref{ap2}) the partial state vector
\begin{equation}
{\bf P} (\vec C')\, |S\rangle \quad \hbox{ with} \quad 
\frac{1}{\langle \Sigma | \Sigma _{N-T} (\vec C') \rangle} \;
|\Sigma _{N-T} (\vec C') \rangle \quad ,
\end{equation}
that is, with its average over the set of basis vectors with clause
vector $\vec C'$ and $N-T$ undetermined variables.
\end{approks}

Following step A, equation (\ref{ap2}) becomes an approximated 
evolution equation for $B$,
\begin{equation} \label{ap3}
B ( \vec C , T+1)  = \sum _{\vec C'}  {\bf \hat H} [ \vec C , \vec C' ;T]
\;  B ( \vec C , T) \quad ,
\end{equation}
where the new evolution matrix ${\bf \hat H}$, not to be confused 
with ${\bf H}$, is
\begin{equation} \label{ap4}
{\bf \hat H} [ \vec C , \vec C' ;T ] =  \frac{ \langle \Sigma (\vec C)|
{\bf H} | \Sigma _{N-T}(\vec C') \rangle} { \langle \Sigma | 
\Sigma _{N-T}(\vec C') \rangle} 
\quad .
\end{equation}
Then,

\begin{approks} Dynamical annealing (step B)

Substitute in equation (\ref{ap3}) the evolution matrix ${\bf \hat H}$ with 
\begin{equation}
{\bf \bar H} [ \vec C , \vec C' ;T] =  \frac{ \overline{\langle \Sigma 
(\vec C)| {\bf H} | \Sigma _{N-T}(\vec C') \rangle}}{\overline{\langle 
\Sigma | \Sigma _{N-T}(\vec C') \rangle} }
\end{equation}
that is, consider the instance ${\bf I}$ is redrawn at each time
step $T\to T+1$, keeping information about clause vectors at time $T$ 
only.
\end{approks}

Let us interpret what we have done so far. The quantity we focus on 
is $\bar B(\vec C;T+1)$, the expectation number of branches at depth
$T$ in the search tree (Figure~\ref{struct}) carrying partial states
with clause vector
$\vec C = (C_1,C_2,C_3)$. Within the dynamical annealing approximation, 
the evolution of the $\bar B$'s is Markovian,
\begin{equation}
\label{bradp}
\bar B(\vec C;T+1)=\sum_{\vec C'}\;{\bf \bar H}\;[\vec C,\vec C';T]\; 
\bar B(\vec C';T) \ .
\end{equation}
The entries of the evolution matrix ${\bf \bar H}[\vec C, \vec C';T]$ 
can be interpreted as the average number of branches with clause vector
$\vec C$ that DPLL will generate through the assignment of one variable
from a partial assignment (partial state) of variables  
with clause vector $\vec C'$.

For the GUC heuristic, we find \citep{Coc},
\begin{eqnarray}
\label{bbra}
&& {\bf\bar H} [\vec C,\vec C'; T] = {C_3' \choose C_3'-C_3} \; \left
(\frac 3{N-T}\right)^{C_3'-C_3} \; \left(1-\frac 3{N-T}\right)^{C_3}
\times \nonumber \\ &&\qquad \qquad \qquad \qquad \sum_{w_2=0}^{C_3'-C_3} \left (\frac
1 2 \right)^{C_3'-C_3} {C_3'-C_3\choose w_2}\times \nonumber \\ 
&& \left \{(1 - \delta _{C'_1} ) \; \left( 1 - \frac1{2(N-T)} \right)^{C'_1-1}
\sum_{z_2=0}^{C_2'} {C_2' \choose z_2} \left (\frac
2{N-T}\right)^{z_2} \right. \times \nonumber \\ &&
 \left (1- \frac 2{N-T}\right)^{C_2'-z_2} 
\sum_{w_1=0}^{z_2} \left (\frac 1 2
\right)^{z_2} {z_2\choose w_1}\; \delta_{C_2-C_2'-w_2+z_2}
\;\delta_{C_1-C_1'-w_1+1}  + \nonumber \\&&
\delta_{C_1'}
\sum_{z_2=0}^{C_2'-1} {C_2'-1 \choose z_2} \left (\frac
2{N-T}\right)^{z_2}\, \left( 1- \frac 2{N-T}\right)^{C_2'-1-z_2}
\times \nonumber \\ && \left.
\sum_{w_1=0}^{z_2} \left(\frac 1 2 \right)^{z_2} {z_2\choose w_1}\;
\delta_{C_2-C_2'-w_2+z_2+1} \; [\delta_{C_1-w_1} + \delta_{C_1-1-w_1}]
 \right\} \ ,
\end{eqnarray}
where $\delta _X$ denotes the Kronecker delta function over integers $X$:
$\delta _X=1$ if $X=0$, $\delta _X=0$ otherwise. Expression (\ref{bbra}) is
easy to obtain from the interpretation following equation (\ref{bradp}).

\subsection{Generating functions and asymptotic scalings at large $N$}

Let us introduce the generating function $G (\,{\vec y}\,;T\,)$ of
the average number of branches $\bar B(\,{\vec C}\,;T\,)$ 
where $\vec y\equiv(y_1,y_2,y_3)$, through
\begin{equation}
\label{gener}
G (\,{\vec y}\,;T\,) =\sum_{\vec C}\; e^{\,{\vec y} \cdot 
{\vec C}}\; \bar B(\,{\vec C}\,,T\,)\quad ,
\quad
{\vec y} \cdot  {\vec C} \equiv \sum _{j=1}^3 y_j \, C_j \quad .
\end{equation}
Evolution equation (\ref{ap3}) for the $\bar B$'s can be rewritten in 
term of the generating function $G$,
\begin{eqnarray}
\label{eqev}
G(\,{\vec y}\,;T+1\,) &=& e^{-\gamma_1({\vec y})} \; G \big(
\,{\vec \gamma}({\vec y})\,; T\,\big) + \nonumber \\
&& \left ( e^{- \gamma_2({\vec y})}
(e^{y_1}+1) - e^{- \gamma_1({\vec y})}\right) \;
G \big(-\infty,\, \gamma_2({\vec y}),\, \gamma_3({\vec y})\,; T\, \big)
\end{eqnarray}
where ${\vec \gamma}$ is a vectorial function of argument ${\vec y}$ whose
components read
\begin{eqnarray}
\gamma_1 ({\vec y}) &=&y_1+\ln\left[1-\frac 1 {2(N-T)} \right]
\quad , \nonumber\\
\gamma_2({\vec y})&=&y_2+\ln\left[1+\frac 2 {N-T} \left (\frac{e^{-y_2}}{2} 
\left(1+ e^{y_1}\right) -1\right)\right]
\quad , \nonumber \\
\gamma_3 ({\vec y})&=&y_3+\ln\left[1+\frac 3 {N-T}  \left (\frac{e^{-y_3}}{2} 
\left(1+ e^{y_2}\right) -1\right)\right] \quad .
\end{eqnarray}
To solve equation (\ref{eqev}), we infer the large $N$ behaviour of 
$G$ from the following remarks:

\begin{enumerate}
\item Each time DPLL assigns
variables through splitting or unit-propagation, the numbers $C_j$ of
clauses of length $j$ undergo $O(1)$ changes. It is thus sensible to
assume that, when the number of assigned variables increases from $T_1 = t\, 
N$ to $T_2=t\, N + \Delta T$ 
with $\Delta T $ very large but $o(N)$ e.g. $\Delta T = \sqrt N$, 
the densities $c_2=C_2/N$ and $c_3=C_3/N$ of 2- and 3-clauses have 
been modified by $o(1)$. 

\item On the same time interval $T_1<T<T_2$, we expect
the number of unit-clauses $C_1$ to vary at each time step. But its
distribution $\rho (C_1|c_2,c_3;t)$, conditioned to the densities 
$c_2$, $c_3$ and the reduced time $t$, should reach some well defined limit 
distribution. This claim is a generalization of the result obtained 
by \citet{Fri} for the analysis of the GUC heuristic in the absence
of backtracking.

\item As long as a partial state does not violate the instance, 
very few unit-clauses
are generated, and splitting frequently occurs. In other words, the
probability that $C_1=0$ is strictly positive as $N$ gets
large. 

\end{enumerate}
 
The above arguments entice us to make the following

\begin{claim} Asymptotic expression for the generating function $G$

For large $N,T$ at fixed ratio $t=T/N$, the generating function 
(\ref{gener}) of the average numbers $\bar B$ of branches 
is expected to behave as
\begin{equation}
\label{scalinghyp2}
G(y_1,y_2, y_3;t\,N ) = \exp \bigg[\, N \, \varphi ( y_2, y_3 ; t)
+ \psi ( y_1,y_2, y_3 ; t) + o(1) \bigg] \quad .
\end{equation}
\end{claim}

Hypothesis (\ref{scalinghyp2}) expresses in a concise way some important
information on the distribution of clause populations 
during the search process that we now extract.
Call $\omega$ the Legendre transform of $\varphi$, 
\begin{equation}
\omega (\,c_2,c_3\,;t\,) = \min _{y_2,y_3} \bigg[ \varphi
(\, y_2,y_3 \, ; t\, ) - y_2\, c_2 - y_3\, c_3\bigg] 
\qquad . \label{inversion}
\end{equation}
Then, combining equations (\ref{gener}), (\ref{scalinghyp2}) and
(\ref{inversion}), we obtain
\begin{equation}
\label{scalinghyp}
\sum _{C_1\ge 0}\rho (C_1|c_2,c_3;t)\; \bar B(C_1,c_2\,N,
c_3\, N; t\, N ) \asymp \exp \big[\,N \, \omega ( c_2,c_3 ; t\,)
\big] \quad ,
\end{equation}
up to non exponential in $N$ corrections. In other words,
the expectation value of the number of branches carrying partial states
with $(1-t)\, N$ undetermined variables and 
$c_j\,N$ $j$-clauses ($j=2,3$) scales exponentially with $N$,
with a growth function $\omega (c_2,c_3;t)$ related to $\varphi (y_2,y_3;t)$ 
through identity (\ref{inversion}).
Moreover, $\varphi (0,0;t)$ is the logarithm of the
number of branches (divided by $N$) after a fraction $t$ of variables
have been assigned. The most probable values of the densities $c_j(t)$
of $j$-clauses are then obtained from the partial derivatives of
$\varphi$: $c_j(t)=\partial \varphi /\partial y_j (0,0)$ for $j=2,3$.

Let us emphasize that $\varphi$ in equation (\ref{scalinghyp2}) does
not depend on $y_1$. This hypothesis simply expresses that, as far as
non violating partial states are concerned,  
both terms on the right hand side of (\ref{eqev}) are of the same order,
and that the density of unit-clauses, $c_1=\partial \varphi/\partial y_1$,
identically vanishes.

Similarly, function $\psi (y_1,y_2,y_3;t)$ is related to the generating
function of distribution $\rho (C_1|c_2,c_3;t)$,
\begin{equation} \label{psigen}
\sum _{C_1\ge 0} \rho (C_1|c_2,c_3;t)\, \e^{y_1\, C_1} =
e^{\psi (y_1,y_2,y_3;t) - \psi(0,y_2,y_3;t)} \quad ,
\end{equation}
where $c_j=\partial \varphi /\partial y_j (y_2,y_3;t)$ ($j=2,3$)
on the left hand side of the above formula.

Inserting expression (\ref{scalinghyp2}) into the evolution equation 
(\ref{eqev}), we find
\begin{eqnarray} \label{mdar}
\frac{\partial \varphi } {\partial t} (y_2,y_3;t) &=& 
-y_1 + \frac 2{1-t} \left[ e^{-y_2} \left( \frac{1+e^{y_1}}2 \right)
-1 \right] \frac{\partial \varphi } {\partial y_2} (y_2,y_3;t) 
\nonumber \\
&+& \frac 3{1-t} \left[ e^{-y_3} \left( \frac{1+e^{y_2}}2 \right)
-1 \right] \frac{\partial \varphi } {\partial y_3} (y_2,y_3;t) 
\nonumber \\
&+& \ln \left[ 1+ K(y_1,y_2) \; e^{\psi (-\infty,y_2,y_3;t)
- \psi(y_1,y_2,y_3;t)} \right]
\end{eqnarray}
where $K(y_1,y_2)=e^{-y_2}(e^{2 \, y_1} + e^{y_1}) -1$. 
As $\varphi$ does not depend upon $y _1$, the latter may be chosen
at our convenience e.g. to cancel $K$ and the contribution from the last 
term in equation (\ref{mdar}),
\begin{equation}
y_1 =
Y_1 (y_2 ) \equiv y_2 - \ln \left(\frac{  1 +\sqrt{1+ 4 e^{y_2}}}2 \right)
\quad .
\end{equation}
Such a procedure, sometimes called kernel method and, to our knowledge, 
first proposed by \citet{knu}, is correct in the major
part of the $y_2,y_3$ space and, in particular, in the vicinity of
$(0,0)$ we focus on in this paper\footnote{It has however to be 
to modified in a small region of the $y_2,y_3$ space; a complete
analysis of this case was carried out by \citet{Coc}.}.
We end up with the following partial differential equation (PDE) for
$\varphi$,
\begin{equation} \label{croi2}
\frac{\partial \varphi } {\partial t} (y_2,y_3;t) = { H} \left[
\frac{\partial \varphi } {\partial y_2} , \frac{\partial \varphi }
{\partial y_3 } , y_2, y_3 , t \right] \quad ,
\end{equation}
where ${H}$ incorporates the details of the splitting 
heuristic\footnote{For the UC heuristic, 
\begin{equation}
{H} _{UC} = {\ln2} +  \frac {3\, c_3}{1-t}\; \left[ e^{-y_3}\left(
\;\frac{1+e^{y_2}}{2}\right) -1 \right]+
\frac{c_2}{1-t}  \; \left(  \frac 32 e^{-y_2}
-2 \right)   \qquad .
\end{equation}},
\begin{eqnarray}
{H} _{GUC} [c_2 ,c_3, y_2, y_3 , t ] &=&  
-Y_1(y_2) +\frac {3\, c_3}{1-t}\; \left[ e^{-y_3}\;\left(
\frac{1+e^{y_2}}{2}\right) -1 \right]\nonumber \\
&+& \frac{c_2}{1-t}  \; \left( e^{ -Y_1(y_2)}
-2 \right)  \qquad .
\end{eqnarray}
We must therefore solve the partial differential equation (PDE) (\ref{croi2})
with the initial condition,
\begin{equation}\label{initphi}
\varphi(y_2,y_3,t=0) = \alpha _0 \; y_3 \quad ,
\end{equation}
obtained through inverse Legendre transform (\ref{inversion}) of the
initial condition over $\bar B$, or equivalently over $\omega$,
\begin{equation}
\omega (c_2,c_3;t=0)= \left\{ \begin{array} {c c c}
0 &\hbox{\rm if} & c_3=\alpha _0\ ,\nonumber \\ -\infty
&\hbox{\rm if} & c_3\ne\alpha _0 \ .\end{array}
\right . \label{condi}
\end{equation}

\begin{center}
\begin{figure}
\includegraphics[height=200pt,angle=-90]{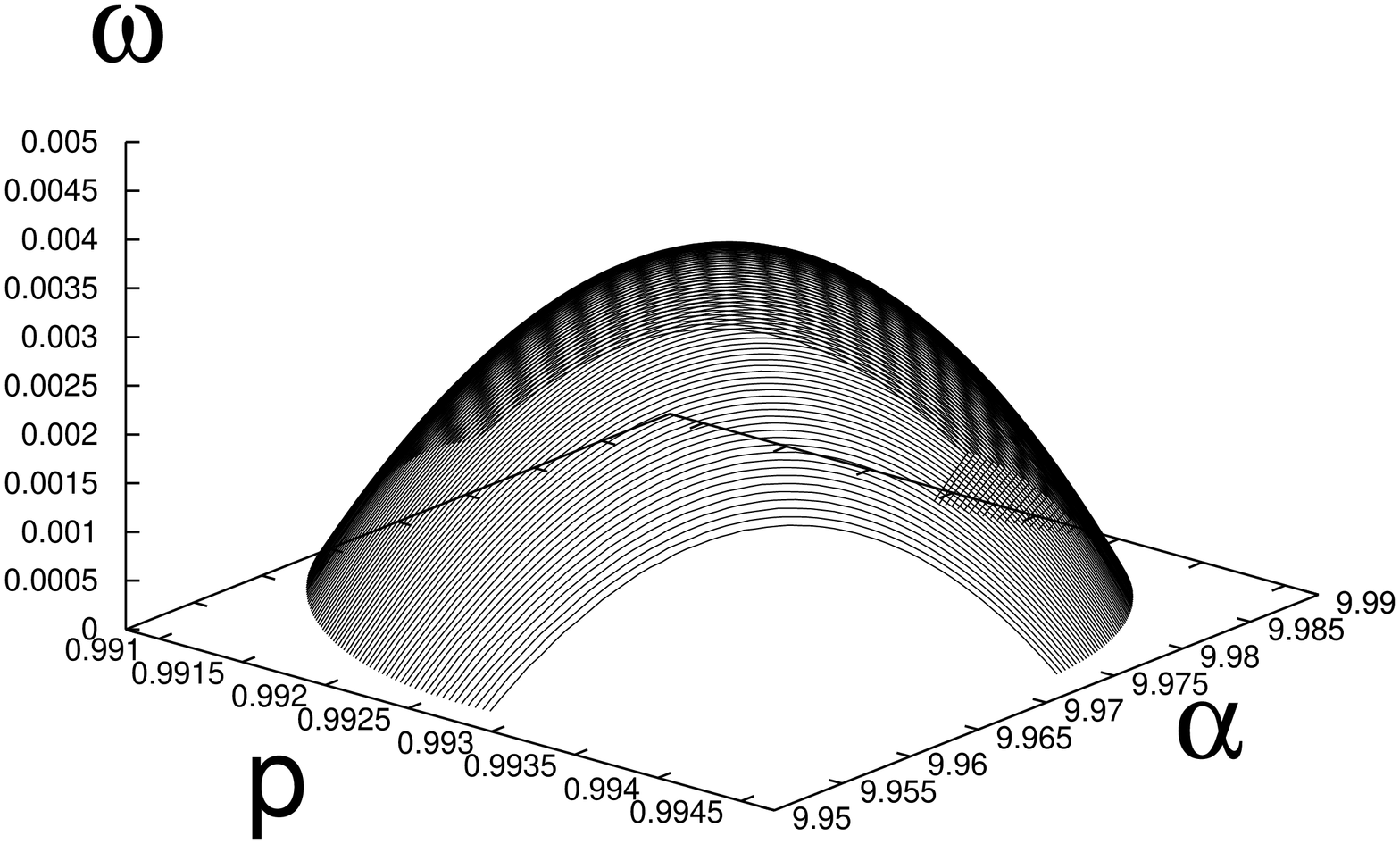}
\includegraphics[height=200pt,angle=-90]{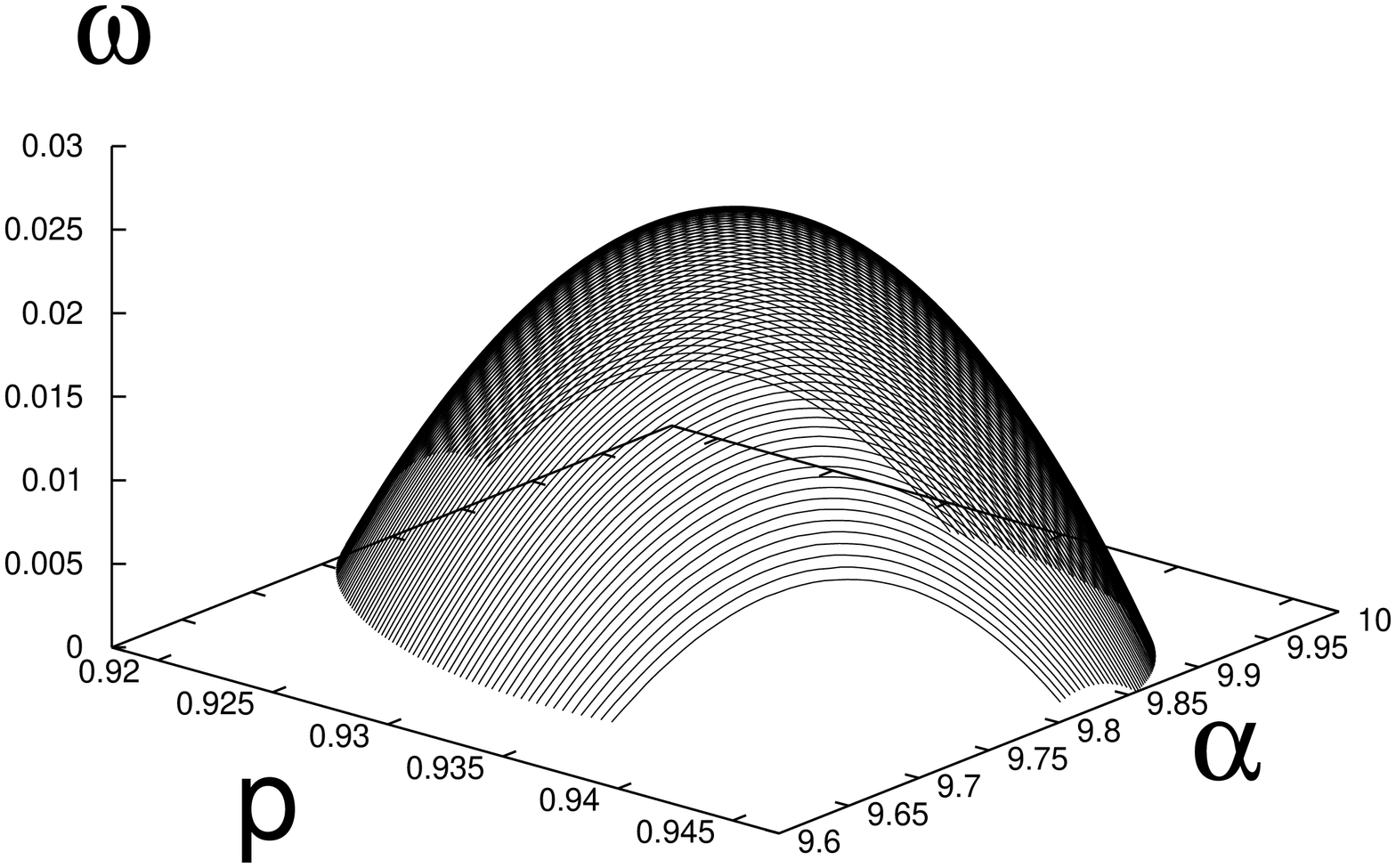} 
$^{\ }$\hskip 3cm \includegraphics[height=200pt,angle=-90]{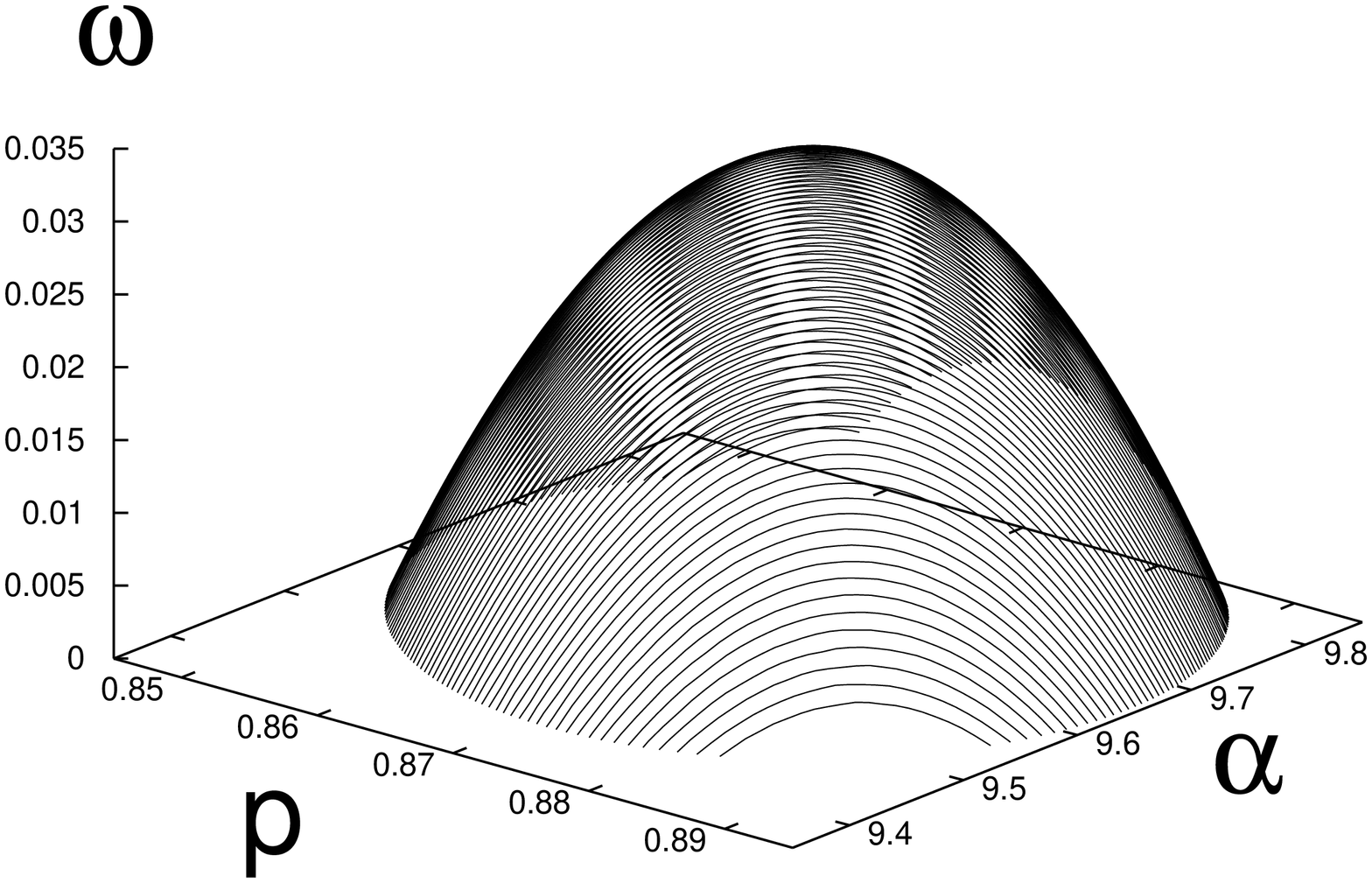}
\vskip .5cm
\caption{Snapshots of the surface $\omega (p,\alpha;t )$ for $\alpha
_0=10$ at three different times {\em i.e.} depths in the tree, 
$t=0.01$, 0.05 and 0.09 (from left to right, top to
down). The height $\omega ^*(t)$ of the top of the surface, with
coordinates $p^*(t), \alpha^*(t)$, is the logarithm (divided by $N$) of the
number of branches. The coordinates $(p^*(t),\alpha ^*(t))$ define the tree
trajectory shown in Figure~\ref{diag}. The halt line is hit at $t_h \simeq 0.094$.
Note the overall growth of the surface $\omega (p,\alpha;t)$ with time
(beware of the change of scales between figures).}
\label{dome}
\end{figure}
\end{center}

\subsection{Interpretation in terms of growth process}

We can interpret the dynamical annealing approximation made in the previous
paragraphs, and the resulting PDE (\ref{croi2}) as a description of 
the growth process of the search tree resulting from DPLL operation.
Using Legendre transform (\ref{inversion}), PDE (\ref{croi2}) can 
be written as an evolution equation for the
logarithm $\omega(c_2,c_3,t)$ of the average number of branches with
parameters $c_2,c_3$ as the depth $t=T/N$ increases,
\begin{equation}
\frac{\partial \omega } {\partial t} (c_2,c_3,t) = { H} \left[ c_2, c_3,
-\frac{\partial \omega } {\partial c_2} , -\frac{\partial \omega }
{\partial c_3 } ,t \right] \qquad . \label{croi}
\end{equation}
Partial differential equation (PDE) (\ref{croi}) is
analogous to growth processes encountered in statistical physics
\citep{Gro}.  The surface $\omega$, growing with ``time'' $t$ above the
plane $c_2,c_3$, or equivalently from (\ref{change}), above the plane 
$p,\alpha$ (Figure~\ref{dome}), describes the whole distribution of branches.
The average number  of branches at depth $t$ in the tree equals
\begin{equation}
B(t) = \int _0 ^1 dp\; \int _0 d\alpha \; e^{N\, \omega (p,\alpha;t)} \asymp
e^{N\, \omega ^* (t)} \quad ,
\end{equation}
where $\omega ^* (t)$ is the maximum over $p,\alpha$ of $\omega (p,\alpha;t)$
reached in $p^*(t), \alpha^*(t)$.
In other words, the exponentially dominant contribution to $B(t)$
comes from branches carrying 2+p-SAT instances with parameters
$p^*(t), \alpha^*(t)$, that is clause densities
$c^*_2(t)= \alpha^*(t) (1-p^*(t))$, $c^*_3(t)= \alpha^*(t) p^*(t)$. 
Parametric plot of $p^*(t),\alpha ^*(t)$ as a function of $t$ 
defines the tree trajectories on Figure~\ref{diag}. 

The hyperbolic line in Figure~\ref{diag} indicates the halt points, where
contradictions prevent dominant branches from further growing.
Each time DPLL assigns a variable through 
unit-propagation, an average number $u(p,\alpha)$ of new 1-clauses is
produced, resulting in a net rate of $u-1$ additional 1-clauses.
As long as $u< 1$, 1-clauses are quickly eliminated and do not
accumulate.  Conversely, if $u  >1$, 1-clauses tend to accumulate. 
Opposite 1-clauses $x$ and $\bar x$ are likely to appear,
leading to a contradiction \citep{Fra,Fri}. The halt line is defined through
$u (p,\alpha)=1$, and reads \citep{Coc},
\begin{equation}
\alpha = \left( \frac{3+\sqrt 5}2 \right) 
\ln \left[ \frac{1+\sqrt 5}2 \right]\;\frac 1{1-p}
\qquad .
\end{equation}
It differs from the halt line $\alpha=1/(1-p)$ corresponding to
a single branch \citep{Fri}. As far as dominant branches are concerned,
an alternative and simpler way of obtaining the halt criterion
is through calculation of the probability
$\rho ^*_S (t) \equiv \rho (C_1=0|c^*_2(t),c^*_3(t);t)$ that a split occurs
when a variable is assigned by DPLL,
\begin{equation}
\rho ^*_S (t) = \exp \left(\frac{\partial \varphi}{\partial t}
(0,0;t)\right) -1  \quad ,
\end{equation}
from equations (\ref{psigen},\ref{mdar}). 
The probability of split vanishes, and unit-clauses accumulate till
a contradiction is obtained, when the tree stops growing. 
Along the tree trajectory, $\omega ^*(t)$ grows thus from 0, on the
right vertical axis, up to some final positive value, $\omega_{THE}$,
on the halt line. $\omega _{THE}$ is our theoretical prediction for the
logarithm of the complexity (divided by $N$)\footnote{Notice that
we have to divide the theoretical value by $\ln 2$ to match the definition
used for numerical experiments; this is done in Table~1}.

Equation (\ref{croi}) was solved using the method of characteristics. 
Using eqn. (\ref{change}), we have plotted the surface $\omega$ at different
times, with the results shown in Figure~\ref{dome} for
$\alpha _0=10$. Values of $\omega _{THE}$, obtained for 
$4.3<\alpha<20$ by solving equation
(\ref{croi}) compare very well with numerical results (Table~1).
We stress that, though our calculation is not rigorous, 
it provides a very good quantitative estimate of the complexity. 
It is therefore expected that our dynamical annealing approximation
be quantitavely accurate. It is a reasonable conjecture that it becomes
exact at large ratios $\alpha_0$, where PDE (\ref{croi2}) can be exactly solved: 

\begin{conj} Asymptotic equivalent of $\omega$ for large ratios

Resolution of PDE (\ref{croi}) in the large ratio $\alpha_0$ limit
gives (for the GUC heuristic),
\begin{equation}
\omega _{THE} (\alpha _0 ) \asymp \frac {3+\sqrt{5}}{6 \,\ln 2}\; 
\left[ \ln \left( \frac{1+\sqrt 5}{2} \right) \right]^2 \; \frac 1{\alpha_0}
\quad .
\end{equation}
This result exhibits the $1/\alpha _0$ scaling proven by \citet{Bea}, and 
is conjectured to be exact.
\end{conj}

As $\alpha _0$ increases, search trees become smaller and smaller, and 
correlations between branches, weaker and weaker, making dynamical
annealing more and more accurate.

\begin{center}
\begin{figure}
\includegraphics[width=400pt,height=250pt,angle=0]{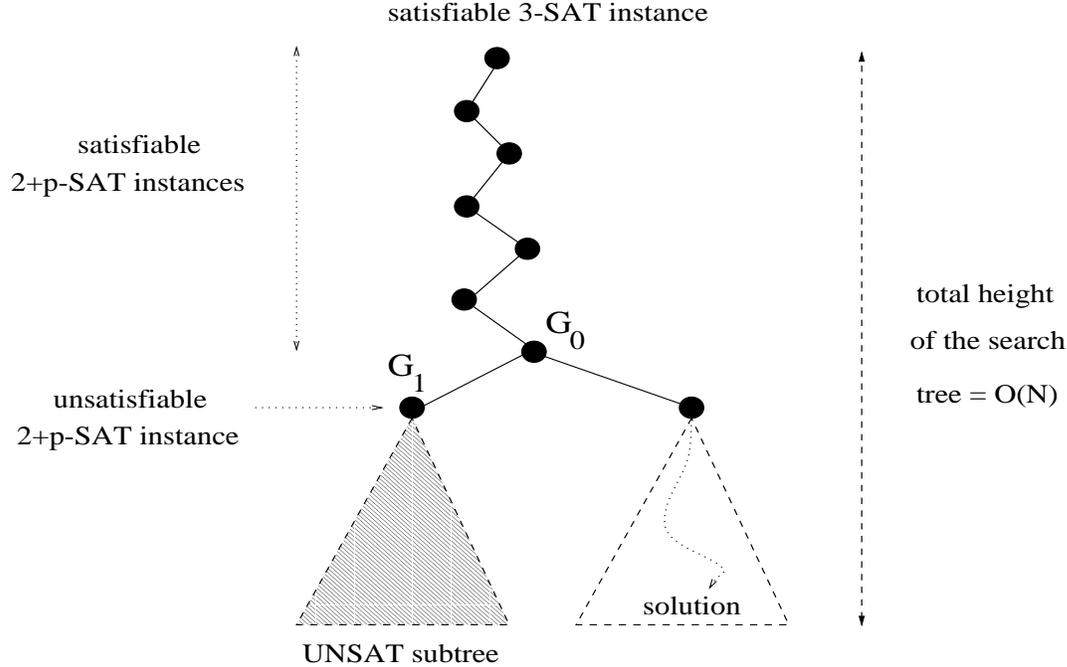}
\vskip .3cm
\caption{Detailed structure of the search tree in the 
upper sat phase ($\alpha _L < \alpha < \alpha _C$). DPLL starts with a 
satisfiable 3-SAT instance and transforms it into a sequence of 
2+p-SAT instances. The leftmost branch in the tree 
symbolizes the first descent made by DPLL. Above node $G_0$, 
instances are satisfiable while below
$G_1$, instances have no solutions. A grey triangle accounts for the 
(exponentially) large refutation subtree that DPLL has to go through
before backtracking above $G_1$ and reaching $G_0$. By definition, the highest
node reached back by DPLL is $G_0$. Further backtracking, below $G_0$, will
be necessary but a solution will be eventually found (right subtree), see
Figure~\ref{trees}C.}
\label{treeinter}
\end{figure}
\end{center}

\section{Upper phase and mixed branch--tree trajectories.}

The interest of the trajectory framework proposed in this paper is best seen
in the upper sat phase, that is, for 
ratios $\alpha _0$ ranging from $\alpha _L$
to $\alpha _C$. This intermediate region juxtaposes branch and tree
behaviors, see search tree in Figures~\ref{trees}C and \ref{treeinter}. 

The branch trajectory,
started from the point $(p=1,\alpha _0)$ corresponding to the initial
3-SAT instance, hits the critical line $\alpha_c(p)$ 
at some point G with coordinates ($p_G,\alpha_G$) after $N\;t_G$ variables 
have been assigned by DPLL, see Figure~\ref{diag}.  The
algorithm then enters the unsat phase and, with high probability,
generates a 2+p-SAT instance
with no solution. A dense subtree that DPLL has to go through
entirely, forms beyond G till the halt line (left subtree in
Figure~\ref{treeinter}).  The size of this subtree
can be analytically predicted from the theory exposed in
Section~3. All calculations are identical, except 
initial condition (\ref{initphi}) which has to be changed into
\begin{equation}
\varphi(y_2,y_3,t=0) = \alpha _G \; (1-p_G) \; y_2 
+\alpha _G \; p_G \; y_3 \quad .
\end{equation} 
As a result we obtain the size $2^{N_G \; \omega _G}$ of the unsatisfiable
subtree to be backtracked (leftmost subtree in Figure~\ref{treeinter}).
$N_G = N\,(1-t_G)$ denotes the number of undetermined variables 
at point $G$. 

$G$ is the highest backtracking node in the tree (Figures~\ref{trees}C
and \ref{treeinter}) reached back by
DPLL, since nodes above G are located in the sat phase and carry
2+p-SAT instances with solutions. DPLL will eventually reach
a solution. The corresponding branch (rightmost
path in Figure~\ref{trees}C) is highly non typical and does not contribute
to the complexity, since almost all branches in the search tree
are described by the tree trajectory issued from G (Figure~\ref{diag}). 
We expect that the computational effort DPLL requires to find
a solution will, to exponential order in $N$, be given by the
size of the left unsatisfiable subtree of Figure~\ref{treeinter}.
In other words, massive backtracking will certainly be present
in the right subtree (the one leading to the solution), and no 
significant statistical difference is expected between both subtrees.

We have experimentally checked this scenario for 
$\alpha _0=3.5$. The average coordinates of the highest
backtracking node, $(p_G\simeq 0.78, \alpha _G \simeq 3.02$), coincide 
with the computed intersection of the single branch trajectory 
(Section 2.2) and the estimated critical line $\alpha_c(p)$ \citep{Coc}. 
As for complexity, experimental measures of $\omega$ from 
3-SAT instances at $\alpha _0= 3.5$, and of $\omega _G$ from
2+0.78-SAT instances at $\alpha _G =3.02$, obey the
expected identity  
\begin{equation}
\omega _{THE} = \omega _G \times (1-t_G) \quad ,
\end{equation} 
and are in very good agreement with theory (Table~1). Therefore, 
the structure of search trees corresponding to instances of 3-SAT 
in the upper sat regime reflects the
existence of a critical line for 2+p-SAT instances.

\section{Conclusions.}

\begin{center}
\begin{figure}
\includegraphics[height=250pt,angle=0]{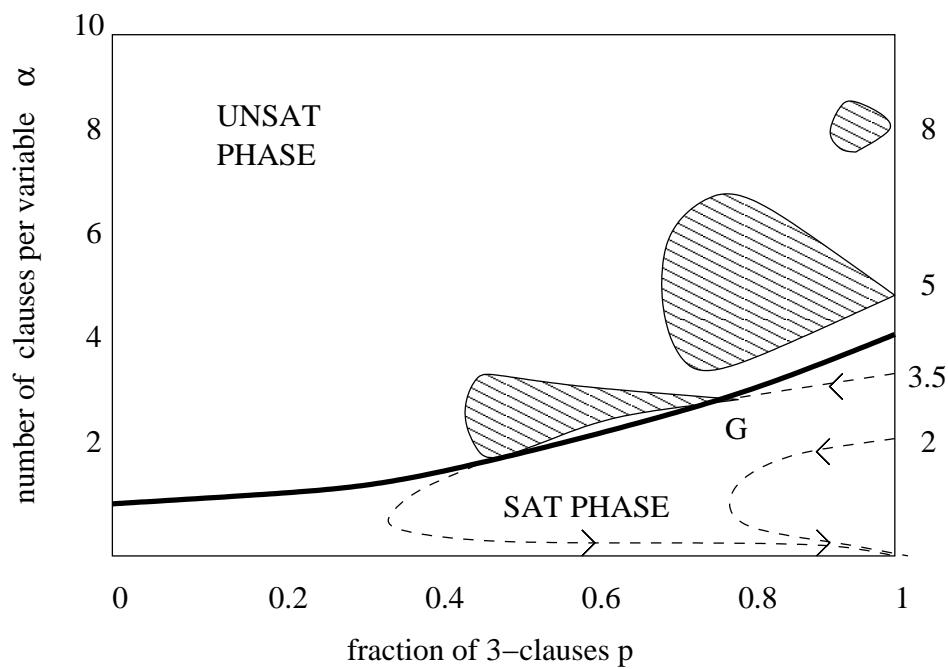}
\caption{Schematic representation of the resolution trajectories in the
sat (branch trajectories symbolized with dashed line) and unsat (tree 
trajectories represented by hatched regions) phases. DPLL goes along
branch trajectories in a linear time, but takes an exponential time
to go through tree trajectories. The mixed case
of hard sat instances correspond to the crossing of the boundary separating
the two phases (bold line), which leads to the exploration of unsat subtrees
before a solution is finally found.}
\label{sche}
\end{figure}
\end{center}

In this paper, we have exposed a procedure to understand the
complexity pattern of the backtrack resolution of the random
Satisfiability problem (Figure~\ref{sche}). Main steps are:
\begin{enumerate}
\item  Identify the  space of parameters in which
the dynamical evolution takes place; this space will be generally larger
than the initial parameter space since the algorithm modifies the instance
structure. While the distribution of 3-SAT instances is characterized by 
the clause per variable ratio 
$\alpha$ only, another parameter $p$ accounting for
the emergence of 2-clauses has to be considered. 
\item Divide the parameter space
into different regions (phases) depending on the output
of the resolution e.g. sat/unsat phases for 2+p-SAT. 
\item Represent the action of  the algorithm as trajectories in this
phase diagram. Intersection of trajectories with the phase boundaries 
allow to distinguish hard from easy regimes (Figure~\ref{sche}). 
\end{enumerate}

In addition, we have also presented 
a non rigorous study of the search tree growth, which allows us to
accurately estimate the complexity of resolution in presence
of massive backtracking. From a mathematical point of view, it is worth 
noticing that monitoring the growth of the search tree requires
a PDE, while ODEs are sufficient to account for the evolution of a single
branch~\citep{Achl}.

An interesting question raised by this picture is the robustness of
the polynomial/exponential crossover point T (Figure 3). While the
ratio $\alpha_L$ separating easy (polynomial) from hard (exponential) 
resolutions depends on the heuristics used by DPLL ($\alpha _L ^{GUC} \simeq
3.003$, $\alpha _L ^{UC} = 8/3$), T appears to be located at the same
coordinates $(p_T=2/5,\alpha _T =5/3)$ for all three UC, GUC, and
SC$_1$ heuristics. From a technical point of view, the robustness of T
comes from the structure of the ODEs (\ref{ode}). The coordinates of
T, and the time $t_T$ at which the branch trajectory issued from
$(p=1,\alpha _0=\alpha _L)$ hits the critical line $\alpha _C(p)$
tangentially, obey the equations $\rho _1 = \partial \rho _1 /\partial
t = 0$ with $\rho _1 = 1- \alpha(t) \big(1-p(t)\big)$.  The set of
ODEs (\ref{ode}), combined with the previous conditions, gives
$p_T=2/5$ \citep{Achl}.

This robustness explains why the polynomial/exponential crossover
location of critically constrained 2+p-SAT instances, which should
{\em a priori} depend  on the algorithm used, was found by
\citet{Sta1} to coincide roughly with the algorithm--independent,
tricritical point on the $\alpha_C(p)$ line.

Our approach has already been extended to other decision problems, e.g. the 
vertex covering of random graphs \citep{Wei} or the coloring
of random graphs \citep{Liat} (see \citep{Cris} for recent rigorous
results on backtracking in this case). It is important to stress that it is
not limited to the determination of the average solving time, but may
also be used to capture its distribution \citep{Gent2,Coc3,Mont} 
and to understand the efficiency of restarts techniques \citep{Gomes}. 
Finally, we emphasize that theorem 6 relates the computational effort
to the evolution operator representing the elementary steps 
of the search heuristic for a {\em given} instance. It is expected
that this approach will be useful to obtain results on the average-case 
complexity of DPLL at fixed instance, where the average is performed over the 
random choices done by the algorithm only \citep{momo}. 

\begin{ack}
We thank J. Franco for his constant support during
the completion of this work. R. Monasson was in part supported by 
the ACI Jeunes Chercheurs ``Algorithmes d'optimisation et
syst\`emes d\'esordonn\'es quantiques'' from the French Ministry of Research.  
\end{ack}

\end{document}